\documentclass[12pt,preprint]{aastex}
\usepackage{emulateapj5}

\shorttitle{ICL}
\shortauthors{Krick, Bernstein}

\begin{document}
\newcommand\msun{\hbox{M$_{\odot}$}}
\newcommand\lsun{\hbox{L$_{\odot}$}}
\newcommand\magarc{mag arcsec$^{-2}$}
\newcommand\h{$h_{70}^{-1}$}

\bibliographystyle{myapj}

\title{\bf Diffuse Optical Light in Galaxy Clusters II: \\ Correlations
with Cluster Properties}

\author{J.E. Krick \altaffilmark{1,2} and R.A. Bernstein \altaffilmark{1}}
\altaffiltext {1}{ Astronomy Department, University of Michigan, Ann Arbor, MI 48109}
\altaffiltext {2}{ Spitzer Science Center, Caltech, Pasadena, CA 91125}
\email{jkrick@caltech.edu, rabernst@umich.edu}

\begin{abstract} 

  We have measured the flux, profile, color, and substructure in the
  diffuse intracluster light (ICL) in a sample of ten galaxy clusters
  with a range of mass, morphology, redshift, and density.  Deep,
  wide-field observations for this project were made in two bands at
  the one meter Swope and 2.5 meter du Pont telescope at Las Campanas
  Observatory.  Careful attention in reduction and analysis was paid
  to the illumination correction, background subtraction, point spread
  function determination, and galaxy subtraction.  ICL flux is
  detected in both bands in all ten clusters ranging from $7.6\times
  10^{10}$ to $7.0\times 10^{11}$ \h \lsun in $r$ and $1.4\times
  10^{10}$ to $1.2\times 10^{11}$ \h \lsun in the $B-$band.  These
  fluxes account for 6 to 22\% of the total cluster light within one
  quarter of the virial radius in $r$ and 4 to 21\% in the $B-$band.
  Average ICL $B-r$ colors range from 1.5 to 2.8 mags when k and
  evolution corrected to the present epoch.  In several clusters we
  also detect ICL in group environments near the cluster center and up
  to ~1\h Mpc distant from the cluster center.  Our sample, having
  been selected from the Abell sample, is incomplete in that it does
  not include high redshift clusters with low density, low flux, or
  low mass, and it does not include low redshift clusters with high
  flux, mass, or density.  This bias makes it difficult to interpret
  correlations between ICL flux and cluster properties.  Despite this
  selection bias, we do find that the presence of a cD galaxy
  corresponds to both centrally concentrated galaxy profiles and
  centrally concentrated ICL profiles.  This is consistent with ICL
  either forming from galaxy interactions at the center, or forming at
  earlier times in groups and later combining in the center.
\end{abstract}

\keywords{galaxies: clusters: individual
  (A4059, A3880, A2734, A2556, A4010, A3888, A3984, A0141, AC114, AC118) ---
  galaxies: evolution --- galaxies: interactions --- galaxies:
  photometry --- cosmology: observations}

\section{Introduction} 

A significant stellar component of galaxy clusters is found outside of
the galaxies.  The standard theory of cluster evolution is one of
hierarchical collapse, as time proceeds, clusters grow in mass through
the merging with other clusters and groups.  These mergers as well as
interactions within groups and within clusters strip stars out of
their progenitor galaxies.  The study of these intracluster stars can
inform hierarchical formation models as well as tell us something
about physical mechanisms involved in galaxy evolution within
clusters.

Paper I of this series \citep{krick2006} discusses the methods of ICL
detection and measurement as well as the results garnered from one
cluster in our sample.  We refer the reader to that paper and the
references therein for a summary of the history and current status of
the field.  This paper presents the remaining nine clusters in the
sample and seeks to answer when and how intracluster stars are formed
by studying the total flux, profile shape, color, and substructure in
the ICL as a function of cluster mass, redshift, morphology, and
density in the sample of 10 clusters.  The advantage to having an
entire sample of clusters is to be able to follow evolution in the ICL
and use that as an indicator of cluster evolution.

Strong evolution in the ICL fraction with mass of the cluster has been
predicted in simulations by both \citet{lin2004} and
\citet{murante2004}. If ongoing stripping processes are dominant, ram
pressure stripping \citep{abadi1999} or harassment \citep{moore1996},
then high mass clusters should have a higher ICL fraction than
low-mass clusters .  If, however, most of the galaxy evolution happens
early on in cluster collapse by galaxy-galaxy merging, then the ICL
should not correlate directly with current cluster mass.

Because an increase in mass is tied to the age of the cluster under
hierarchical formation, evolution has also been predicted in the ICL
fraction as a function of redshift \citep{willman2004,rudick2006}.
Again, if ICL formation is an ongoing process then high redshift
clusters will have a lower ICL fraction than low redshift clusters.
Conversely, if ICL formation happened early on in cluster formation
there will be no correlation of ICL with redshift.

The stripping of stars (or even the gas to make stars) to create an
intracluster stellar population requires an interaction between their
original host galaxy and either another galaxy, the cluster potential,
or possibly the hot gas in the cluster.  Because all of these
processes require an interaction, we expect cluster density to be a
predictor of ICL fraction.  Cluster density is linked to cluster
morphology, which implies morphology should also be a predictor of ICL
fraction.  Specifically we measure morphology by the presence or
absence of a cD galaxy.  cD galaxies are the results of 2 - 5 times
more mergers than the average cluster galaxy \citep{dubinski1998}.
The added number of interactions that went into forming the cD galaxy
will also mean an increased disruption rate in galaxies therefore
morphological relaxed (dynamically old) clusters should have a higher
ICL flux than dynamically young clusters.  

Observations of the color and fractional flux in the ICL over a sample
of clusters with varying redshift and dynamical state will allow us to
identify the timescales involved in ICL formation.  If the ICL is the
same color as the cluster galaxies, it is likely to be a remnant from
ongoing interactions in the cluster.  If the ICL is redder than the
galaxies it is likely to have been stripped from galaxies at early
times.  Stripped stars will passively evolve toward red colors while
the galaxies continue to form stars.  If the ICL is bluer than the
galaxies, then some recent star formation has made its way into the
ICL, either from ellipticals with low metallicity or spirals with
younger stellar populations, or from in situ formation.

While multiple mechanisms are likely to play a role in the complicated
process of formation and evolution of clusters, important constraints
can come from ICL measurement in clusters with a wide range of
properties.  In addition to directly constraining galaxy evolution
mechanisms, the ICL flux and color is a testable prediction of
cosmological models.  As such it can indirectly be used to examine the
accuracy of the physical inputs to these models.

%------------------------

This paper is structured in the following manner.  In \S2 we discuss
the characteristics of the entire sample.  Details of the observations
and reduction are presented in \S3 and \S4 including flat-fielding,
sky background subtraction methods, object detection, and object
removal and masking.  In \S5 we lists the results for both cluster and
ICL properties including a discussion of each individual
cluster. Accuracy limits are discussed in \S6.  A discussion of the
interesting correlations can be found in \S7 followed by a summary of
the conclusions in \S8.

Throughout this paper we use $H_0=70$km/s/Mpc, $\Omega_M$ = 0.3,
$\Omega_\Lambda$ = 0.7.

%%%%%%%%%%%%%%%%%%%%%%%%%%%%%%%%%%%%%%%%%%%%%%%%%%%%%%5

\section{The Sample}
\label{sample2}

The general properties of our sample of ten galaxy clusters have been
outlined in paper I; for completeness we summarize them briefly here.
Our choice of the 10 clusters both minimizes the observational hazards
of the galactic and ecliptic plane, and maximizes the amount of
information in the literature.  All clusters were chosen to have
published X--ray luminosities which guarantees the presence of a
cluster and provides an estimate of the cluster's mass.  The ten
chosen clusters are representative of a wide range in cluster
characteristics, namely redshift ($0.05 < z < 0.3$), morphology (3
with no clear central dominant galaxy, and 7 with a central dominant
galaxy as determined from this survey, \S \ref{morphology2}, and not
from Bautz Morgan morphological classifications), spatial projected density
(richness class 0 - 3), and X--ray luminosity ($1.9\times10^{44}$
ergs/s $< Lx < 22\times10^{44}$ ergs/s).  We discuss results from the
literature and this survey for each individual cluster in order of
ascending redshift in \S \ref{results2}.

\section{Observations}
\label{observations2}

The sample is divided into a ``low'' ($0.05<z<0.1$) and ``high''
($0.15 < z< 0.3$) redshift range which we have observed with the 1
meter Swope and 2.5 meter du Pont telescope respectively.  The du Pont
observations were discussed in detail in paper I.  The Swope
observations follow a similar observational strategy and data
reduction process which we outline below.  Observational parameters
are listed in Table \ref{tab:obspars2}.

We used the $2048\times3150$ ``Site\#5'' CCD with a $3 e^-/$count gain
and $7e^-$ readnoise on the Swope telescope.  The pixel scale is
0.435\arcsec/pixel ($15\mu$ pixels), so that the full field of view
per exposure is $14.8\arcmin \times 22.8\arcmin$.  Data was taken in
two filters, Gunn-$r$ ($\lambda_0 = 6550$ \AA) and $B$ ($\lambda_0 =
4300$ \AA). These filters were selected to provide some color
constraint on the stellar populations in the ICL by spanning the
4000\AA\ break at the relevant redshifts, while avoiding flat-fielding
difficulties at longer wavelengths and prohibitive sky brightness at
shorter wavelengths.

Observing runs occurred on October 20-26, 1998, September 2-11, 1999,
and September 19-30, 2000.  All observing runs took place within eight
days of new moon.  A majority of the data were taken under photometric
conditions.  Those images taken under non-photometric conditions were
individually tied to the photometric data (see discussion in
\S\ref{reduction2}.  Across all three runs, each cluster was observed
for an average of 5 hours in each band. In addition to the cluster
frames, night sky flats were obtained in nearby, off-cluster,
``blank'' regions of the sky with total exposure times roughly equal
to one third of the integration times on cluster targets.  Night sky
flats were taken in all moon conditions.  Typical $B-$ and $r-$band
sky levels during the run were $22.7$ and $21.0$ \magarc,
respectively.

Cluster images were dithered by one third of the field of view between
exposures.  The large overlap from the dithering pattern gives us
ample area for linking background values from the neighboring cluster
images.  Observing the cluster in multiple positions on the chip
reduces large-scale flat-fielding fluctuations upon combination.
Integration times were typically 900 seconds in $r$ and 1200 seconds
in $B$.

%---------------------------------------------------------------------------------

\section{Reduction}
\label{reduction2}

In order to create mosaiced images of the clusters with a uniform
background level and accurate resolved--source fluxes, the images were
bias and dark subtracted, flat--fielded, flux calibrated,
background--subtracted, extinction corrected, and registered before
combining. Methods for this are discussed in detail in paper I and
summarized below.

%---------------------------------------------------------------------------------

%\subsection{Bias and Dark Subtraction}

The bias level is roughly 270 counts which changed by approximately
8\% throughout the night.  This, along with the large-scale ramping
effect in the first 500 columns of every row was removed in the
standard manner using IRAF tasks.  The mean dark level is 1.6
counts/900s, and there is some vertical structure in the dark which
amounts to 1.4 counts/900s over the whole image.  To remove this
large-scale structure from the data images, a combined dark frame from
the whole run was median smoothed over $9\times9$ pixels
($3.9\arcsec$), scaled by the exposure time, and subtracted from the
program frames.  Small scale variations were not present in the dark.
%\subsection{Flat Fielding}
Pixel--to--pixel sensitivity variations were corrected in all cluster
and night-sky flat images using nightly, high S/N, median-combined
dome flats with 70,000 -- 90,000 total counts.  After this step, a
large-scale illumination pattern remains across the chip.  This was
removed using night-sky flats of ``blank'' regions of the sky, which,
when combined using masking and rejection, produced an image with no
evident residual flux from sources but has the large scale
illumination pattern intact.  The illumination pattern was stable
among images taken during the same moon phase.  Program images were
corrected only with night sky flats taken in conditions of similar
moon.

%---------------------------------------------------------------------------------

%\subsection{Non-linearity}

We find that the Site\#3 CCD does have an approximately $7\%$
non-linearity over the full range of counts, which we fit with a
second order polynomial and corrected for in all the data.  The same
functional fit was found for both the 1998 and 1999 data, and also
applied to the 2000 data.  The uncertainty in the linearity correction
is incorporated in the total photometric uncertainty.

%---------------------------------------------------------------------------------

%\subsection{Photometric Calibration}
%\label{photometry}

Photometric calibration was performed in the usual manner using
Landolt standards at a range of airmasses.  Extinction was monitored
on stars in repeat cluster images throughout the night.  Photometric
nights were analyzed together; solutions were found in each filter for
an extinction coefficient and common magnitude zero-point with a $r-$
and $B-$band {\sc RMS} of 0.04 \& 0.03 magnitudes in October 1998,
0.03 \& 0.03 magnitudes in September 1999, and 0.05 \& 0.05 magnitudes
in September 2000.  These uncertainties are a small contribution to
our final error budget (\S\ref{noise2}).  Those exposures taken in
non-photometric conditions were individually tied to the photometric
data using roughly 10 stars well distributed around each frame to find
the effective extinction for that frame.  Among those non-photometric
images we find a standard deviation of 0.03 magnitudes within each
frame.  Two further problems with using non-photometric data for low
surface brightness measurements are the scattering of light off of
clouds causing a changing background illumination across the field and
secondly the smoothing out of the PSF.  We find no spatial gradient
over the individual frame to the limit discussed in \S \ref{noise2}.
The change in PSF is on small scales and will have no effect on the
ICL measurement (see \ref{star2}).

%-------------------------------------------------------------------------------

%\subsection{Sky Background Subtraction}

Due to the temporal variations in the background, it is necessary to
link the off-cluster backgrounds from adjacent frames to create one
single background of zero counts for the entire cluster mosaic before
averaging together frames. To determine the background on each
individual frame we measure average counts in approximately twenty
$20\times20$ pixel regions across the frame.  Regions are chosen
individually by hand to be a representative sample of all areas of the
frame that are more distant than $0.8 h_{70}^{-1}$Mpc from the center
of the cluster.  This is well beyond the radius at which ICL
components have been identified in other clusters \citep{krick2006,
  feldmeier2002, gonzalez2005, zibetti2005}.  The average of these background
regions for each frame is subtracted from the data, bringing every
frame to a zero background.  The accuracy of the background
subtraction is discussed in \S \ref{noise2}.

%---------------------------------------------------------------------------------

%\subsection{Extinction Correction}

The remaining flux in the cluster images after background subtraction
is corrected for atmospheric extinction by multiplying each individual
image by $10^{\tau\chi/2.5}$, where $\chi$ is the airmass and $\tau$
is the fitted extinction in magnitudes from the photometric solution.
This multiplicative correction is between 1.04 and 2.0 for an airmass
range of 1.04 to 1.9.

%$$ m = m_{o} - 2.5\log(F) -\tau\chi  $$

%---------------------------------------------------------------------------------

%\subsection{Registration \& Distortion}

The IRAF tasks {\sc geomap} and {\sc geotran} were used to find and
apply x and y shifts and rotations between all images of a single
cluster.  The {\sc geotran} solution is accurate on average to 0.03
pixels (RMS).  Details of the final combined image after
pre-processing, background subtraction, extinction correction, and
registration are included in Table \ref{tab:obspars2}.

%---------------------------------------------------------------------------------
\subsection{Object Detection}
\label{detection2}

Object detection follows the same methods as Paper I.  We use
SExtractor to both find all objects in the combined frames, and to
determine their shape parameters.  The detection threshold in the $V$,
$B$, and $r$ images was defined such that objects have a minimum of 6
contiguous pixels, each of which are greater than $1.5\sigma$ above
the background sky level.  We choose these parameters as a compromise
between detecting faint objects in high signal-to-noise regions and
rejecting noise fluctuations in low signal-to-noise regions.  This
corresponds to minimum surface brightnesses which range from of 25.2
to 25.8 \magarc\ in $B$, 25.9 to 26.9 \magarc\ in $V$, and 24.7 to
26.4 \magarc\ in $r$ (see Table \ref{tab:obspars2}).  This range in
surface brightness is due to varying cumulative exposure time in the
combined frames.  Shape parameters are determined in SExtractor using
only those pixels above the detection threshold.

%---------------------------------------------------------------------------------

\subsection{Object Removal \& Masking}
\label{remove_obj2}

To measure the ICL we remove all detected objects from the frame by
either subtraction of an analytical profile or masking. Details of
this process are described below.

%---------------------------------------------------------------------------------

\subsubsection{Stars}
\label{star2}
\begin{figure}
%\centering{\psfig{file=f2.jpg,width=3in}}
\centering
%%\epsscale{.45}
%\includegraphics{f1.jpg}
\includegraphics[scale=0.45]{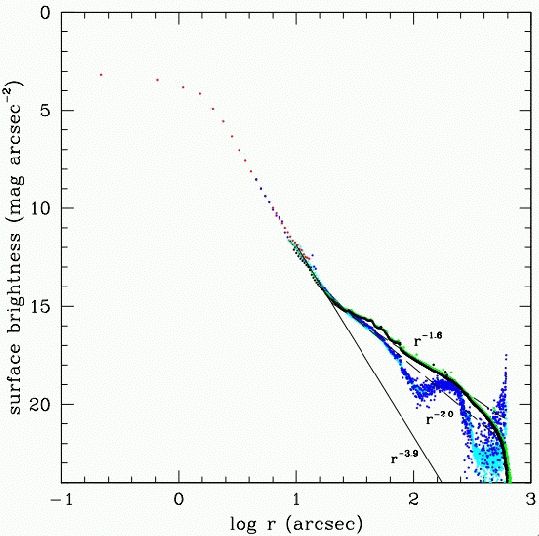}
\caption[The PSF of the 40-inch Swope telescope at Las Campanas
Observatory]{The PSF of the 40-inch Swope telescope at Las Campanas
Observatory.  The y-axis shows surface brightness scaled to correspond
to the total flux of a zero magnitude star.  The profile within
5\arcsec\ was measured from unsaturated stars and can be affected by
seeing. The outer profile was measured from two stars with
super-saturated cores imaged in two different bands.  The profile with
the bump in it at 100\arcsec\ is the $r-$band profile, that without
the bump is the $B-$band PSF.  The bump in the profile at 100\arcsec\
is due to a reflection off the CCD which then bounces off of the
filter, and back down onto the CCD.  The outer surface brightness
profile decreases as $r^{-2}$ in the $r-$band and $r^{-1.6}$ in the
$B$, shown by the dashed lines.  An $r^{-3.9}$ profile is plotted to
show the range in slopes.}
\label{fig:psf2}
\end{figure}

Scattered light in the telescope and atmosphere produce an extended
point spread function (PSF) for all objects.  To correct for this
effect, we determine the extended PSF using the profiles of a
collection of stars from super-saturated 4th mag stars to unsaturated
14th magnitude stars.  The radial profiles of these stars were fit
together to form one PSF such that the extremely saturated star was
used to create the profile at large radii and the unsaturated stars
were used for the inner portion of the profile.  This allows us to
create an accurate PSF to a radius of $7$\arcmin, shown in Figure
\ref{fig:psf2}.

The inner region of the PSF is well fit by a Moffat function.  The
outer region is well fit by $r^{-2.0}$ in the $r-$band and $r^{-1.6}$
in the $B-$band.  In the $r-$band there is a small additional halo of
light at roughly 50 - 100\arcsec (200-400pix) around stars imaged on
the CCD.  The newer, higher quality, anti-reflection coated
interference $B-$band filter does not show this halo, which implies
that the halo is caused by reflections in the filter.  To test the
effect of clouds on the shape of the PSF we create a second deep PSF
from stars in cluster fields taken under non-photometric conditions.
There is a slight shift of flux in the inner 10 arcseconds of the PSF
profile, which will have no impact on our ICL measurement.

For each individual, non-saturated star, we subtract a scaled
band--specific profile from the frame in addition to masking the inner
$30$\arcsec\ of the profile (the region which follows a Moffat
profile).  For each individual saturated star, to be as cautious as
possible with the PSF wings, we have subtracted a stellar profile
given the USNO magnitude of that star, and produced a large mask to
cover the inner regions and any bleeding.  The mask size is chosen to
be twice the radius at which the star goes below 30\magarc, and
therefore goes well beyond the surface brightness limit at which we
measure the ICL.  We can afford to be liberal with our saturated star
masking since most clusters have very few saturated stars which are
not near the center of the cluster where we need the unmasked area to
measure any possible ICL.

In the specific case of A3880 there are two saturated stars (9th and
10th $r-$band magnitude) within two arcminutes of the core region of
the cluster.  If we used the same method of conservatively masking
(twice the radius of the 30 \magarc\ aperture), the entire central
region of the image where we expect to find ICL would be lost.  We
therefore consider a less extreme method of removing the stellar
profile by iteratively matching the saturated stars' profiles with the
known PSF shape.  We measure the saturated star profiles on an image
which has had every object except for those two saturated stars
masked, as described in \S \ref{galaxies2}.  We can then scale our
measured PSF to the star's profile, at radii where there is expected
to be no contamination from the ICL, and the star's flux is not
saturated.  Since the two stars are within an arcminute of each other,
the scaled profiles of the stars are iteratively subtracted from the
masked cluster image until the process converges on solutions for the
scaling of each star.  We still use a mask for the inner region ($\sim
75\arcsec$) where saturation and seeing effect the profile shape.

%---------------------------------------------------------------------------------

\subsubsection{Galaxies}
\label{galaxies2}

We want to remove all the flux in our images associated with galaxies.
Although some galaxies might follow deVaucouleurs, Sersic, or
exponential profiles, those galaxies which are near the centers of
clusters can not be fit with these or other models either because of
the overcrowding in the center or because their profiles really are
different due to their location in a dense environment.  A variety of
strategies for modeling galaxies within the centers of clusters were
explored in Paper 1 and were found to be inadequate for these
purposes.  Since we can not fit and subtract the galaxies to remove
their light, we instead mask all galaxies in our cluster images.

By masking, we remove from our ICL measurements all pixels above a
surface brightness limit which are centered on a galaxy as detected by
SExtractor.  For paper I, we chose to mask inside of 2 - 2.3 times the
radius at which the galaxy light dropped below 26.4 \magarc\ in $r$,
akin to 2-2.3 times a Holmberg radius \citep{holmberg1958}.  Holmberg
radii are typically used to denote the outermost radii of the stellar
populations in galaxies.

Galaxy profiles will also have the characteristic underlying shape of
the PSF, including the extended halo.  However for a 20th magnitude
galaxy, the PSF is below 30 \magarc by a radius of 10\arcsec.

Each of the clusters has a different native surface brightness
detection threshold based on the illumination correction and
background subtraction, and they are all at different redshifts.
However we want to mask galaxies at all redshifts to the same physical
surface brightness to allow for a meaningful comparison between
clusters at different redshifts.  To do this we make a correction for
$(1+z)^4$ surface brightness dimming and a $k$ correction for each
cluster when calculating mask sizes.  The masks sizes change by an
average of 10\% and at most 22\% from what they would have been given
the native detection threshold. Both the native and corrected surface
brightness detection thresholds are listed in Table
\ref{tab:obspars2}. To test the effect of mask size on the ICL profile
and total flux, we also create masks which are $30\%$ larger and
$30\%$ smaller in area than the calculated mask size.  The flux within
the masked areas for these galaxies is on average 25\% more than the
flux identified by SExtractor as the corrected isophotal magnitude for
each object.

%------------------------------------------------------------------------------\
\section{Results}
\label{results2}
Here we discuss our methods for measuring both cluster and ICL
properties as well as a discussion of each individual cluster in our
sample.

\subsection{Cluster Properties}

Cluster redshift, mass, and velocity dispersion are taken from the
literature, where available, as listed in table
\ref{tab:characteristics2}.  Additional properties that can be
identified in our data, particularly those which may correlate with
ICL properties (cluster membership, flux, dynamical state, and global
density), are discussed below and also summarized in Table
\ref{tab:characteristics2}.

\subsubsection{Cluster Membership \& Flux}
\label{member2}

\begin{figure}
%%\epsscale{.45}
%centering
\includegraphics[scale=0.45]{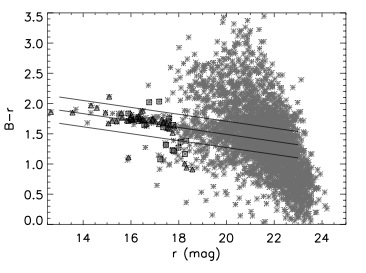}
\includegraphics[scale=0.45]{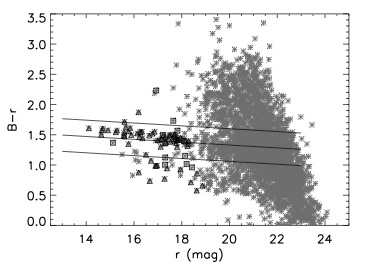}
\includegraphics[scale=0.45]{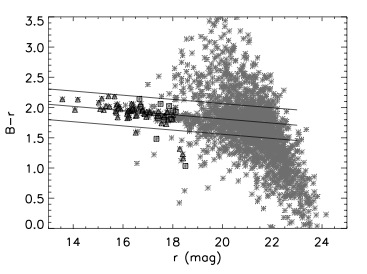}
\includegraphics[scale=0.45]{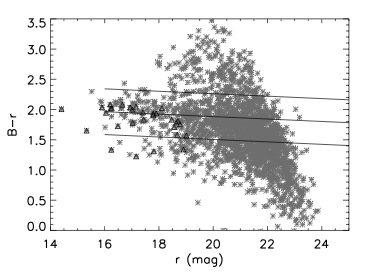}
\includegraphics[scale=0.45]{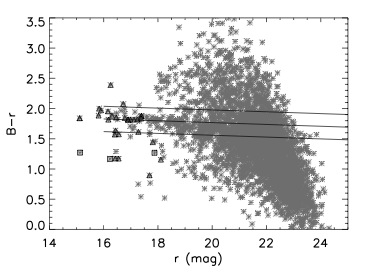}
\includegraphics[scale=0.45]{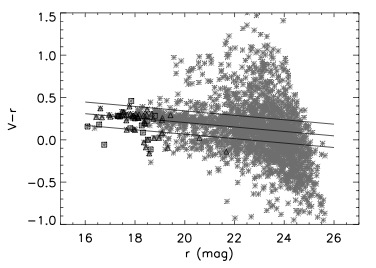}
\includegraphics[scale=0.45]{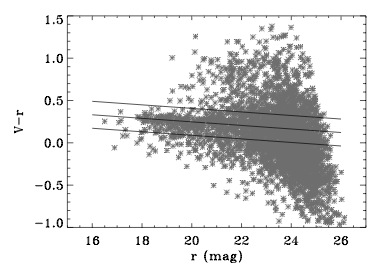}
\includegraphics[scale=0.45]{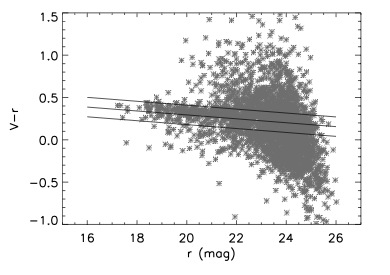}
\includegraphics[scale=0.45]{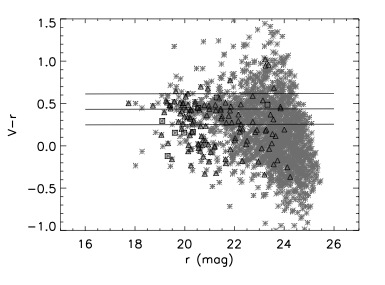}
\includegraphics[scale=0.45]{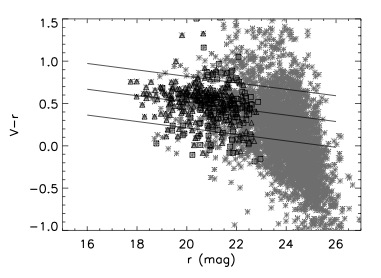}
\caption[allcmd]{The color magnitude diagrams for all ten clusters in
  increasing redshift order from let to right, top to bottom; A4059,
  A3880, A2734, A2556, A4010, A3888, A3984, A014, AC114, AC118.  All
  galaxies detected in our image are denoted with a gray star.  Those
  galaxies which have membership information in the literature are
  over--plotted with open black triangles (members) or squares
  (non--members)(membership references are given in \S \ref{A4059}-
  \ref{AC118}).  Solid lines indicate a biweight fit to the red
  sequence with $1\sigma$ uncertainties. }
\label{fig:allcmd}
%\epsscale{1}
\end{figure}

Cluster membership and galaxy flux are both determined using a color
magnitude diagram (CMD) of either $B - r$ vs. $r$ (clusters with $z <
0.1$) or $V - r$ vs. $r$ (clusters with $z > 0.1$).  We create color
magnitude diagrams for all clusters using corrected isophotal
magnitudes as determined by SExtractor.  Membership is then assigned
based on a galaxy's position in the diagram.  If a given galaxy is
within $1 \sigma$ of the red cluster sequence (RCS) determined with a
biweight fit, then it is considered a member (fits are shown in Figure
\ref{fig:allcmd}).  All others are considered to be
non-member foreground or background galaxies.  This method selects the
red elliptical galaxies as members.  The benefits of this method are
that membership can easily be calculated with 2 band photometry.  The
drawbacks are that it both does not include some of the bluer true
members and does include some of the redder non-members.  An
alternative method of determining cluster flux without spectroscopy by
integrating under a background subtracted luminosity function is
discussed in detail in \S5.3 of paper I.  Due to the large
uncertainties involved in both methods ($\sim30\%$), the choice of
procedure will not greatly effect the conclusions. 

To determine the total flux in galaxies, we sum the flux of all member
galaxies within the same cluster radius.  The image size of our
low-redshift clusters restricts that radius to one quarter of the
virial radius of the cluster where virial radii are taken from the
literature or calculated from X--ray temperatures as described in
\S\ref{A4059}-\ref{AC118}.  From tests with those clusters where we do
have some spectroscopic membership information from the literature
(see \S\ref{A2734} \& \S\ref{A38882}), we expect the uncertainty in
flux from using the CMD for membership to be $\sim30\%$.

Fits to the CMDs produce the mean color of the red ellipticals, the
slope of the color versus magnitude relation (CMR) for each cluster,
and the width of that distribution.  Among our 10 clusters, the color
of the red sequence is correlated with redshift whereas the slopes of
the relations are roughly the same across redshift, consistent with
\citet{lopezcruz2004}.  The widths of the CMRs vary from 0.1 to 0.4
magnitudes.  This is expected if these clusters are made up of
multiple clumps of galaxies all at similar, but not exactly the same,
redshifts.  True background and foreground groups and clusters can
also add to the width of the RCS.

In order to compare fluxes from all clusters, we consider two
correction factors.  First, galaxies below the detection threshold
will not be counted in the cluster flux as we have measured it, and
will instead contribute to the ICL flux.  Since each cluster has a
different detection threshold based mainly on the quality of the
illumination correction (see Table \ref{tab:obspars2}), we calculate
individually for each cluster the flux contribution from galaxies
below the detection threshold.  Without luminosity functions for each
cluster, we adopt the \citet{goto2002} luminosity function based on
200 Sloan clusters ($\alpha_r'=-0.85\pm0.03$). The flux from dwarf
galaxies below the detection threshold ( $M=-11$ in $r$) is less than
or equal to 0.1\% of the flux from sources above the detection
threshold (our assumed value of total flux).  This is an extremely
small contribution due to the faint end slope, and our deep, uniform
images with detection thresholds in all cases more than 7 magnitudes
dimmer than $M_*$.  Our surface brightness detection thresholds are
low enough that we don't expect to miss galaxies of normal surface
brightness below our detection threshold at any redshift assuming that
all galaxies at all redshifts have similar central surface
brightnesses. 

Second, we apply k and evolutionary corrections to account for the
shifting of the bandpasses through which we are observing, and the
evolution of the galaxy spectra due to the range in redshifts we
observe.  We use \citet{poggianti1997} for both of these corrections
as calculated for simple stellar population of elliptical galaxies in
$B$, $V$, and $r$.

\subsubsection{Dynamical Age}
\label{morphology2}

Dynamical age is an important cluster characteristic for this work as
dynamical age is tied to the number of past interactions among the
galaxies.  We discuss four methods for estimating cluster dynamical
age based on optical and X--ray imaging. The first two methods are
based on cluster morphology using Bautz Morgan type and an indication
of the presence of a cD galaxy.  We use morphology as a proxy for
dynamical age since clusters with single large elliptical galaxies at
their centers (cD) have presumably been through more mergers and
interactions than clusters that have multiple clumps of galaxies where
none have settled to the center of the potential.  Those clusters with
more mergers are dynamically older, therefore clusters with cD
galaxies should be dynamically older.  Specifically Bautz Morgan type is a
measure of cluster morphology defined such that type I clusters have
cD galaxies, type III clusters do not have cD galaxies, and type II
clusters may show cD-like galaxies which are not centrally located.
Bautz Morgan type is not reliable as Abell did not have membership
information.  To this we add our own binary indicator of cluster
morphology; clusters which have single galaxy peaks in the centers of
their ICL distributions (cD galaxies) versus clusters which have
multiple galaxy peaks in the centers of their ICL distributions (no
cD).

We have more information about the dynamical age of the cluster beyond
just the presence or absence of a cD galaxy, namely the difference in
brightness of the brightest cluster galaxy (BCG) relative to the next
few brightest galaxies in the cluster \citep[the luminosity gap
statistic][]{milosavljevic2005}, which is our third estimate of
dynamical age.  Clusters with one bright galaxy that is much brighter
than any of the other cluster galaxies imply an old dynamic age
because it takes time to form that bright galaxy through multiple
mergers.  Conversely, multiple evenly bright galaxies imply a cluster
that is dynamically young.  For our sample we measure the magnitude
differences between the first (M1) and second (M2) brightest galaxies
that are considered members based on color.  We run the additional
test of comparing M2-M1 with M3-M1, where consistency between these
values insures a lack of foreground or background contamination.
Values of M3-M1 range from 0.24 to 1.1 magnitudes and are listed in
Table \ref{tab:characteristics2}.  This is the most reliable measure
of dynamic age available to us in this dataset. In a sample of 12
galaxy groups from N-body hydrodynamical simulations,
\citet{donghia2005} find a clear, strong correlation between the
luminosity gap statistic and formation time of the group (spearman
rank coefficient of 0.91) such that $\delta$mag increases by $0.69 \pm
0.41(1\sigma)$ magnitudes for every one billion years of formation.
We assume this simulation is also an accurate reflection of the evolution
of clusters and therefore that M3-M1 is well correlated with formation
time and therefore dynamical age of the clusters.

The fourth method for measuring dynamical state is based on the X--ray
observations of the clusters.  In a simulation of 9 cluster mergers
with mass ratios ranging from 1:1 to 10:1 with a range of orbital
properties, \citet{poole2006} show that clusters are virialized at or
shortly after they visually appear relaxed through the absence of
structures (clumps, shocks, cavities) or centroid shifts (X--ray peak
vs. center of the X--ray gas distribution).  We then assume that
spherically distributed hot gas as evidenced by the X--ray
morphologies of the clusters free from those structures and centroid
shifts implies relaxed clusters which are therefore dynamically older
clusters that have already been through significant mergers.  With
enough photons, X--ray spectroscopy can trace the metallicity of
different populations to determine progenitor groups or clusters.
X--ray observations are summarized in \S \ref{A4059} - \S \ref{AC118}.

\subsubsection{Global Density}

Current global cluster density is an important cluster characteristic for this
work as density is correlated with the past interaction rate among
galaxies.  We would like a measure of the number of galaxies in each
of the clusters within some well defined radius which encompasses the
potentially dynamically active regions of the cluster.  Abell chose to
calculate global density as the number of galaxies with magnitudes
between that of the third ranked member, M3, and M3+2 within 1.5
Mpc of the cluster, statistically correcting for foreground and
background galaxy contamination with galaxy densities outside of
1.5Mpc \citep{abell1989}. The cluster galaxy densities are then binned
into richness classes with values of zero to three, where richness
three clusters are higher density than richness zero clusters.
Cluster richnesses are listed in Table
\ref{tab:characteristics2}.

In addition to richness class we use a measure of global density which
has not been binned into coarse values and is not affected by sample
completeness.  To do this we count the number of member galaxies
inside of 0.8 \h Mpc to the same absolute magnitude limit for all
clusters.  Membership is assigned to those galaxies within $1\sigma$
of the color magnitude relation (CMR).  The density may be affected by
the width of the CMR if the CMR has been artificially widened due to
foreground and background contamination.  We choose a magnitude limit
of $M_{r}$ = -18.5 which is deep enough to get many tens of galaxies
at all clusters, but is shallow enough that our photometry is still
complete.  At the most distant clusters (z=0.31), an $M_{r}$ = -18.5
galaxy is a $125\sigma$ detection.  The numbers of galaxies in each
cluster that meet these criteria range from 62 - 288, and are in good
agreement with the broader Abell richness determination.  These
density estimates are listed in Table \ref{tab:characteristics2}.

%------------------------------------------------------------------------------
\subsection{ICL properties}

We detect an ICL component in all ten clusters of our sample.  We
describe below our methods for measuring the surface brightness
profile, color, flux, and substructure in that component.

\subsubsection{Surface brightness profile}
\label{profile2}

In eight out of 10 clusters the ICL component is centralized enough to
fit with a single set of elliptical isophotes.  The exceptions are
A0141 and AC118.  We use the IRAF routine {\sc ellipse} to fit
isophotes to the diffuse light which gives us a surface brightness
profile as a function of semi--major axis.  The masked pixels are
completely excluded from the fits.  There are 3 free parameters in the
isophote fitting: center, position angle (PA), and ellipticity.  We
fix the center and let the PA and ellipticity vary as a function of
radius.  Average ICL ellipticities range from 0.3 to 0.7 and vary
smoothly if at all within each cluster.  The PA is notably coincident
with that of the cD galaxy where present (discussed in \S \ref{A4059}
- \ref{AC118}).

We identify the surface brightness profile of the total cluster light
(ie., including resolved galaxies) for comparison with the ICL within
the same radial extent. To do this, we make a new ``cluster'' image by
masking non-member galaxies as determined from the color magnitude
relation (\S \ref{member2}).  A surface brightness profile of the
cluster light is then measured from this image using the same
elliptical isophotes as were used in the ICL profile measurement.

Figure \ref{fig:predict} shows the surface brightness profiles of all
eight clusters for which we can measure an ICL profile.  Individual
ICL profiles in both $r-$ and $V-$ or $B-$bands are shown in Figures
\ref{fig:A4059} - \ref{fig:A118}.  Results based on all three versions
of mask size (as discussed in \S \ref{galaxies2}) are shown via
shading on those plots.  Note that we are not able to directly measure
the ICL at small radii ($< \sim 70kpc$) in any of the clusters because
greater than 75\% of those pixels are masked.  The uncertainty in the
ICL surface brightness is dominated by the accuracy with which the
background level can be identified, while the error on the mean within
each elliptical isophote is negligible, as discussed in \S
\ref{noise2}.  Error bars in Figures \ref{fig:predict} and
\ref{fig:A4059} - \ref{fig:A118} show the $1\sigma$ uncertainty based
on the error budget for each cluster (see representative error budget in Table
\ref{tab:error2}).

The ICL surface brightness profiles have two interesting
characteristics.  First, in all cases they can be fit by both
exponential and deVaucouleurs profiles.  Both appear to perform
equally well given the large error bars at low surface
brightness. These profiles, in contrast to the galaxy profiles, are
relatively smooth, only occasionally reflecting the clustering of
galaxies. Second, the ICL is more concentrated than the galaxies,
which is to say that the ICL falls off more rapidly with increased
radius than the galaxy light.  In all cases the ICL light is
decreasing rapidly enough at large radii such that the additional flux
beyond the radius at which we can reliable measure the surface
brightness is at most 10\% of the flux inside of that radius based on
an extrapolation of the exponential fit.

There are 2 clusters (A0141, Figure \ref{fig:A141} \& AC118, Figure
\ref{fig:A118}) for which there is no single centralized ICL profile.
These clusters do not have a cD galaxy, and their giant ellipticals
are distant enough from each other that the ICL is not a continuous
centralized structure.  We therefore have no surface brightness
profile for those clusters although we are still able to measure an
ICL flux, as discussed below.

We attempt to measure the profile of the cD galaxy where present in
our sample. To do this we remove the mask of that galaxy and allow
ellipse to fit isophotes all the way into the center.  In 5 out of 7
clusters with a cD galaxy, the density of galaxies at the center is so
great that just removing the mask for the cD galaxy is not enough to
reveal the center of the cluster due to the other overlapping
galaxies.  Only for A4059 and A2734 are we able to connect the ICL
profile to the cD profile at small radii.  These are shown in Figures
\ref{fig:A4059} \& \ref{fig:A2734}.

In both cases the entire profile of the cD plus ICL is well fit by a
single DeVaucouleurs profile, although it can also be fit by 2
DeVaucouleurs profiles.  The profiles can not be fit with single
exponential functions.  We do not see a break between the cD and ICL
profiles as seen by \citet{gonzalez2005}.  While those authors find
that breaks in the extended BCG profile are common in their sample,
$\sim 25\%$ of the BCG's in that sample did not show a clear
preference for a double deVaucouleurs model over the single
deVaucouleurs model.  In both clusters where we measure a cD profile,
the color appears to start out with a blue color gradient, and then
turn around and become increasingly redder at large radii as the ICL
component becomes dominant (see Figures \ref{fig:A4059} \&
\ref{fig:A2734}).

\subsubsection{ ICL Flux }

The total amount of light in the ICL and the ratio of ICL flux to
total cluster flux can help constrain the importance of galaxy
disruption in the evolution of clusters.  As some clusters have cD
galaxies in the centers of their ICL distribution, we need a
consistent, physically motivated method of measuring ICL flux in the
centers of those clusters as compared to the clusters without a single
centralized galaxy.  The key difference here is that in cD clusters
the ICL stars will blend smoothly into the galaxy occupying the center
of the potential well, whereas with non-cD clusters the ICL stars in
the center are unambiguous.  Since our physical motivation is to
understand galaxy interactions, we consider ICL to be all stars which
were at some point stripped from their original host galaxies,
regardless of where they are now.

In the case of clusters with cD galaxies, although we cannot separate
the ICL from the galaxy flux in the center of the cluster, we can
measure the ICL profile outside of the cD galaxy.
\citet{gonzalez2005} have shown for a sample of 24 clusters that a BCG
with ICL halo can be well fit with two deVaucouleurs profiles.  The
two profiles imply two populations of stars which follow different
orbits. We assume stars on the inner profile are cD galaxy stars and
those stars on the outer profile are ICL stars.  \citet{gonzalez2005}
find that the outer profile on average accounts for 80\% of the
combined flux and becomes dominant at 40-100kpc from the center which
is at surface brightness levels of 24 - 25 \magarc\ in $r$.  Since all
of our profiles are well beyond this radius and well below this
surface brightness level, we conclude that the ICL profile we identify
is not contaminated by cD galaxy stars.  Assuming that the stars on
the outer profile have different orbits than the stars on the inner
profile, we calculate ICL flux by summing all the light in the outer
profile from a radius of zero to the radius at which the ICL becomes
undetectable.  Note that this method identifies ICL stars regardless
of their current state as bound or unbound from the cD galaxy.

We therefore calculate ICL flux by first finding the mean surface
brightness in each elliptical annuli where all masked pixels are not
included.  This mean flux is then summed over all pixels within that
annulus including the ones which were masked.  This represents a
difference from paper I where we performed an integration on the fit
to the ICL profile; here we sum the profile values themselves.  We are
justified in using the area under the galaxy masks for the ICL sum
since the galaxies only account for less than 3\% of the volume of the
cluster regardless of projected area.

There are two non-cD clusters (A141 \& A118) for which we could not
recover a profile.  We calculate ICL flux for those clusters by
measuring a mean flux within three concentric, manually--placed,
elliptical annuli (again not utilizing masked pixels) in the mean, and
then summing that flux over all pixels in those annuli.  All ICL
fluxes are subject to the same k and evolutionary corrections as
discussed in \S \ref{member2}.

\subsubsection{ ICL Fraction }
\label{iclfraction}
In addition to fluxes, we present the ratio of ICL flux to total
cluster flux, where total cluster flux includes ICL plus galaxy flux.
Galaxy flux is taken from the CMDs out to $0.25r_{virial}$, as
discussed in \S \ref{member2}.  ICL fractions range from 6 to 22\% in
the $r-$band and 4 to 21\% in the $B-$band where the smallest fraction
comes from A2556 and the largest from A4059.  All fluxes and fractions
are listed in Table \ref{tab:characteristics2}.  As mentioned in \S
\ref{member2}, there is no perfect way of measuring cluster flux
without a complete spectroscopic survey.  Based on those clusters
where we do have some spectroscopic information, we estimate the
uncertainty in the cluster flux to be $\sim30\%$ .  This includes both
the absence from the calculation of true member galaxies, and the
false inclusion of non-member galaxies.  

All cluster fluxes as measured from the RCS do not include blue member
galaxies so those fluxes are potentially lower limits to the true
cluster flux, implying that the ICL fractions are potentially biased
high.  This possible bias is made more complicated by the known fact
that not all clusters have the same amount of blue member galaxies
\citep{butcher1984}.  Less evolved clusters (at higher redshifts) will
have higher fractions of blue galaxies than more evolved clusters (at
lower redshifts).  Therefore ICL fractions in the higher redshift
clusters will be systematically higher than in the lower redshift
clusters since their fluxes will be systematically underestimated.  We
estimate the impact of this effect using blue fractions from
\citet{couch1998} who find maximal blue fractions of $60\%$ of all
cluster galaxies at $z=0.3$ as compared to $\sim20\%$ at the present
epoch.  If none of those blue galaxies were included in our flux
measurement for AC114 and AC118 (the two highest z clusters), this
implies a drop in ICL fraction of $\sim 40\%$ as compared to $\sim
10\%$ at the lowest redshifts.  This effect will strengthen the
relations discussed below.

Most simulations use a theoretically motivated definition of ICL which
determine its fractional flux within $r_{200}$ or $r_{vir}$.  It is
not straightforward to compare our data to those simulated values
since our images do not extend to the virial radius nor do they extend
to infinitely low surface brightness which keeps us from measuring
both galaxy and ICL flux at those large radii. The change in
fractional flux from $0.25r_{vir}$ to $r_{vir}$ will be related to the
relative slopes of the galaxies versus ICL.  As the ICL is more
centrally concentrated than the galaxies we expect the fractional flux
to decrease from $0.25r_{vir}$ to $r_{vir}$ since the galaxies will
contribute an ever larger fraction to the total cluster flux at large
radii.  We estimate what the fraction at $r_{vir}$ would be for 2
clusters in our sample, A4059 and A3984 (steep profile and shallow
profile respectively), by extrapolating the exponential fits to both the ICL and
galaxy profiles.  Using the extrapolated flux values, the fractional
flux decreases by 10\% where ICL and galaxy profiles are steep and up
to 90\% where profiles are shallower.

%---------------------------------------------------------------------

\subsubsection{Color}

For those clusters with an ICL surface brightness profile we measure a
color profile as a function of radius by binning together three to
four points from the surface brightness profile.  All colors are k
corrected and evolution corrected assuming a simple stellar population
\citep{poggianti1997}.  Color profiles range from flat to increasingly
red or increasingly blue color gradients (see Figures
\ref{fig:allcolor}).  We fit simple linear functions to the color
profiles with their corresponding errors.  To determine if the color
gradients are statistically significant we look at the $\pm2\sigma$
values on the slope of the linear fit.  If those values do not include
zero slope, then we assume the color gradient is real.  Color error
bars are quite large, so in most cases $2\sigma$ does include a flat
profile.  The significant color gradients (A4010, A3888, A3984) are
discussed in \S \ref{A4059} - \ref{AC118}.

For all clusters an average ICL color is used to compare with cluster
properties.  In the case where there is a color gradient, that average
color is taken as an average of all points with error bars less than
one magnitude.

%---------------------------------------------------------------------------------

\subsubsection{ICL Substructure}
\label{iclsub2}

Using the technique of unsharp masking (subtracting a smoothed version
of the image from itself) we scan each cluster for low surface
brightness (LSB) tidal features as evidence of ongoing galaxy
interactions and thus possible ongoing contribution to the ICL .  All
10 clusters do have multiple LSB features which are likely from tidal
interactions between galaxies, although some are possibly LSB galaxies
seen edge on.  For example we see multiple interacting galaxies and
warped galaxies, as well as one shell galaxy.  For further discussion
see \S6.5 of paper I.  From the literature we know that the two
highest redshift clusters in the sample (AC114 and AC118, z=0.31) have
a higher fraction of interacting galaxies than other clusters
\citep[$\sim 12\%$ of galaxies, ][]{couch1998}.  In two of our
clusters, A3984 and A141, there appears to be plume-like structure in
the diffuse ICL, which is to say that the ICL stretches from the BCG
towards another set of galaxies.  Of this sample, only A3888 has a
large, hundred kpc scale, arc type feature, see Figure 9 and Table 2
of paper I.  There are $\sim 4$ examples of these large features in
the literature \citep{gregg1998,
  calcaneo2000,feldmeier2004,mihos2005}.  These structures are not
expected to last longer than a few cluster crossing times, so we don't
expect that they must exist in our sample.  Furthermore, it is
possible that there is significant ICL substructure below our surface
brightness limits \citep{rudick2006}.

%-------------------------------------------------------------------------
\subsubsection{Groups}
\label{groups2}
In seven out of 10 clusters the diffuse ICL is determined by eye to be
multi-peaked \\(A4059,A2734,A3888,A3984,A141,AC114,AC118).  In some
cases those excesses surround the clumps of galaxies which appear to
all be part of the same cluster, ie the clumps are within a few
hundred kpc from the center but have obvious separations, and there is
no central dominant galaxy (eg., A118). In other cases, the secondary
diffuse components are at least a Mpc from the cluster center (eg.,
A3888).  In these cases, the secondary diffuse light component is
likely associated with groups of galaxies which are falling in toward
the center of the cluster, and may be at various different stages of
merging at the center.  This is strong evidence for ICL creation in
group environments, which is consistent with recent measurements of a
small amount of ICL in isolated galaxy groups \citep{castro2003,
  durrell2004, rocha2005}.  This is also consistent with current
simulations \citep[][and references therein]{willman2004, fujita2004,
  gnedin2003a, rudick2006, sommer-larsen2006}. From the theory, we
expect ICL formation to be linked with the number density of galaxies.
Since group environments can have high densities at their centers and
have lower velocity dispersions, it is not surprising that groups have
ICL flux associated with them.  \citet{sommer-larsen2006} find the
intra-group light to have very similar properties to the ICL making up
$12-45\%$ of the group light, having roughly deVaucouleurs profiles,
and in general varying in flux from group to group where groups with
older dynamic ages \citep[fossil groups][]{donghia2005} have a larger
amount of ICL.  Groups in individual clusters are discussed in
\S\ref{A4059} - \ref{AC118}.

%--------------------------------------------------------------------------------
\subsection{Accuracy Limits}
\label{noise2}

The accuracy of the ICL surface brightness is limited on small scales
($<10\arcsec$) by photon noise.  On larger scales ($>10\arcsec$),
structure in the background level (be it intrinsic or instrumental)
will dominate the error budget.  We determine the stability of the
background level in each cluster image on large scales by first median
smoothing the masked image by 20\arcsec.  We then measure the mean
flux in thousands of random 1\arcsec\ regions more distant than 0.8
Mpc from the center of the cluster.  The standard deviation of these
regions represents the accuracy with which we can measure the
background on $20\arcsec$ scales.  We tested the accuracy of this
measure for even larger-scale uncertainties on two clusters (A3880
from the 40'' data and A3888 from the 100'' data).  We find that
the uncertainty remains roughly constant on scales equal to, or larger
than, $20\arcsec$.  These accuracies are listed for each cluster in
Table \ref{tab:obspars2}.  Regions from all around the frame are used
to check that this estimate of standard deviation is universal across
the image and not affected by location in the frame.  This empirical
measurement of the large-scale fluctuations across the image is
dominated by the instrumental flat-fielding accuracy, but includes
contributions from the bias and dark subtraction, physical variations
in the sky level, and the statistical uncertainties mentioned above.
 
We examine the effect of including data taken under non-photometric
conditions on the large-scale background illumination.  This noise is
fully accounted for in the measurement described above.  All $B-$ and
$V-$ band data were taken on photometric nights.  Five clusters
include varying fractions of non-photometric $r-$ band data; $47\%$ of
A3880, $12\%$ of A3888, $15\%$ of A3984, $48\%$ of A141, and $14\%$ of
A114 are non-photometric.  For A3880, the cluster with one of the
largest fractions of non-photometric data, we compare the measured
accuracy on the combined image which includes the non-photometric data
with accuracy measured from a combined image which includes only
photometric frames.  The resulting large-scale accuracy is 0.3
\magarc better on the frame which includes only photometric data.
Although this does imply that the non-photometric frames are noisier,
the added signal strength gained from having 4.5 more hours on source
outweighs the extra noise.

This empirical measurement of the large--scale background fluctuations
is likely to be a conservative estimate of the accuracy with which we
can measure surface brightness on large scales because it is derived
from the outer regions of the image where compared to the central
regions on average a factor of $\sim 2$ fewer individual exposures
have been combined for the 100'' data and a factor of $~1.5$ for the
40'' (which has a larger field of view and requires less dithering).
A larger number of dithered exposures at a range of airmass, lunar
phase, photometric conditions, time of year, time of night, and
distance to the moon has the effect of smoothing out large-scale
fluctuations in the illumination pattern.  We therefore expect greater
accuracy in the center of the image where the ICL is being measured.

We include a list all sources of uncertainty for one cluster in our
sample (A3888) in Table \ref{tab:error2} (reproduced here from Paper
I).  In addition to the dominant uncertainty due to the large-scale
fluctuations on the background as discussed above, we quantify the
contributions from the photometry, masking, and the accuracy with
which we can measure the mean in the individual elliptical isophotes.
Errors for the other clusters are similarly dominated by background
fluctuations, which are listed in Table \ref{tab:obspars2}.  The errors
on the total ICL fluxes in all bands range from 17\% to 70\% with an
average of 39\%.  The exception is A2556 which reaches a flux error of
100\% in the $B-$band due to its extremely faint profile (see \S
\ref{A2556}).  Assuming a 30\% error in the galaxy flux (see \S
\ref{member2}), the errors on the ICL fraction are on average 48\%.
The errors plotted on the surface brightness profiles are the
$1\sigma$ errors.

%-----------------------------------------------------------------------------
\section{Discussion}
\label{discuss2}

We measure a diffuse intracluster component in all ten clusters in our
sample.  Clues to the physical mechanisms driving galaxy evolution
come from comparing ICL properties with cluster properties.  We have
searched for correlations between the entire set of properties.  Pairs
of properties not explicitly discussed below showed no correlations.
Limited by a small sample and non-parametric data, we use a Spearman
rank test to determine the strength of any possible correlations where
1.0 or -1.0 indicate a definite correlation or anti--correlation
respectively, and 0 indicates no correlation.  Note that this test
does not take into account the errors in the parameters, and instead
only depends on their rank among the sample.  Where a correlation is
indicated we show the fit as well as $\pm2\sigma$ in both y-intercept
and slope to graphically show the ranges of the fit, and give some
estimate of the strength of the correlation.

There are selection biases in our data between cluster parameters
due to our use of an Abell selected sample.  The Abell
cluster sample is incomplete at high redshifts; it does not include
low-mass, low-luminosity, low-density, high-redshift clusters because
of the difficulty in obtaining the required sensitivity with
increasing redshift.  Although our 5 low-redshift clusters are not
affected by this selection effect, and should be a random sampling,
small numbers prevent those clusters from being fully representative
of the entire range of cluster properties.

Specifically we discuss the possibility that there is a real trend
underlying the selection bias in the cases of lower luminosity (Figure
\ref{fig:clusterz}) and lower density clusters (Figure
\ref{fig:ngalsz}) being preferentially found at lower redshift.
Clusters in our sample with less total galaxy flux are preferentially
found at low redshifts, however hierarchical formation predicts the
opposite trend; clusters should be gaining mass over time and hence
light over time.  Note that on size scales much larger than the virial
radius mass does not change with time and therefore those systems can
be considered as closed boxes; but on the size scales of our data, a
quarter of a virial radius, clusters are not closed boxes.

We might expect a slight trend, as was found, such that lower density
clusters are found at lower redshifts. As a cluster ages, it
converts a larger number of galaxies into a smaller number of galaxies
via merging and therefore has a lower density at lower redshifts
despite being more massive than high redshift clusters.  The infall of
galaxies works against this trend.  The sum total of merger and infall
rates will control this evolution of density with redshift.  The
observed density redshift relation for this sample is strong; over the
range z=0.3 - 0.05 (elapsed time of 3Gyr assuming standard $\Lambda
CDM$) the projected number density of galaxies has to change by a
factor of 5.5, implying that every 5.5 galaxies in the cluster must
have merged into 1 galaxy in the last 3 Gyr.  This is well above a
realistic merger rate for this timescale and this time period
\citep{gnedin2003b}.  Instead it is likely that we are seeing the
result of a selection effect.

An interesting correlation which may be indirectly due to the
selection bias is that clusters with less total galaxy flux tend to
have lower densities (Figure \ref{fig:ngalsicl}).  While we expect a
smaller number of average galaxies to emit a smaller amount of total
light, it is possible that the low density clusters are actually made
up of a few very bright galaxies.  So although the trend might be
real, it is also likely that the redshift selection effect of both
density and cluster flux is causing these two parameters to be
correlated.

A correlation which does not appear to be affected by sample
selection is that lower density clusters in our sample are weakly
correlated with the presence of a cD galaxy, see Figure
\ref{fig:ngalsm3m1}. A possible explanation for this is that as a
cluster ages it will have made a cD galaxy out of many smaller
galaxies, so the density will actually be lower for dynamically older
clusters.  \citet{loh2006} find the same correlation by looking at a
sample of environments around 2000 SDSS luminous red galaxies.

In the remainder of this section we examine the interesting physics
that can be gleaned from the combination of cluster properties and ICL
properties given the above biases.  The interpretation of ICL
correlations with cluster properties is highly complicated due not
only to small number statistics and the selection bias, but to the
direction of the selection bias.  Biases in mass, density, and total
galaxy flux with redshift will destructively combine to cancel the
trends which we expect to find in the ICL (as described in the
introduction).  An added level of complication is due to the fact that
we expect the ICL flux to be evolving with time.  We examine below
each ICL property in turn, including how the selection bias will
effect any conclusions drawn from the observed trends.
%-------------

\subsection{ICL flux}
We see a range in ICL flux likely caused by the differing interaction
rates and therefore differing production of tidal tails, streams,
plumes, etc. in different clusters.  Clusters include a large amount
of tidal features at low surface brightness as evidenced by their
discovery at low redshift where they are not as affected by surface
brightness dimming \citep{mihos2005}.  It is therefore not surprising
that we see a variation of flux levels in our own sample.

ICL flux is apparently correlated with three cluster parameters;
M3-M1, density, and total galaxy flux (Figures \ref{fig:m3m1icl},
\ref{fig:ngalsicl}, \& \ref{fig:clustericl}).  There is no direct,
significant correlation between ICL flux and redshift.  As discussed
above, the selection effects of density and mass with redshift will
tend to cancel any expected trends in either density, mass, or
redshift.  We therefore are unable to draw conclusions from these
correlations.  \citet{zibetti2005}, who have a sample of 680 SDSS
clusters, are able to split their sample on both richness and
magnitude of the BCG (as a proxy for mass). They find that both richer
clusters and brighter BCG clusters have brighter ICL than poor or
faint clusters.

%-----------------------------
\subsubsection{ICL Flux vs. M3-M1}
\label{m3m1}
Figure \ref{fig:m3m1icl} shows the moderate correlation between ICL
flux and M3-M1 such that clusters with cD galaxies have less ICL than
clusters without cD galaxies (Spearman coefficient of -0.50). Although
we choose M3-M1 to be cautious about interlopers, M2-M1 shows the same
trend with a slightly more significant spearman coefficient of -0.61.
Our simple binary indicator of the presence of a cD galaxy gives the
same result.  Clusters with cD galaxies (7) have an average flux of
$2.3\pm0.96\times10^{11} (1\sigma)$ whereas clusters without cD
galaxies (3) have an average flux of $5.0\pm0.18\times10^{11}
(1\sigma)$.

Although density is correlated with M3-M1, and density is affected by
incompleteness, this trend of ICL flux with M3-M1 is not necessarily
caused by that selection effect.  Furthermore, the correlation of
M3-M1 with redshift is much weaker (if there at all) than trends of
either density or cluster flux with redshift.  If the observed
relation is due to the selection effect then we are prevented from
drawing conclusions from this relation.  Otherwise, if this relation
between ICL flux and the presence of a cD galaxy is not caused by a
selection effect, then we conclude that the lower levels of measured
ICL are a result of the ICL stars being indistinguishable form the cD
galaxy and therefore the ICL is evolving in a similar way to a cD
galaxy.

By which physical mechanism can the ICL stars
end up in the center of the cluster and therefore overlap with cD
stars?  cD galaxies indicate multiple major mergers of galaxies which
have lost enough energy or angular momentum to now reside in the
center of the cluster potential well. ICL stars on their own will not
be able to migrate to the center over any physically reasonable
timescales unless they were stripped at the center, or are formed in
groups and get pulled into the center along with their original
groups\citep{merritt1984}.

Assuming the ICL is observationally inseparable from the cD galaxy, we
investigate how much ICL light the measured relation implies is hidden
amongst the stars of the cD galaxy.  If 20\% of the total cD + ICL
light is added to the value of the ICL flux in the outer profile, then
the observed trend of ICL flux with M3-M1 is weakened (Spearman
coefficient drops from 0.5 to 0.4).  If 30\% of the total cD + ICL
light is hidden in the inner profile then the relation disappears
(Spearman coefficient of 0.22).  The measured relation between ICL
$r-$band flux and dynamical age of the clusters may then imply that
25-40\% of the ICL is coincident with the cD galaxy in dynamically
relaxed clusters.

%--------------------------------------------------------
\subsection{ICL fraction}
\label{iclfrac}
We focus now on the fraction of total cluster light which is in the
diffuse ICL.  If ICL and galaxy flux do scale together (not just due
to the selection effect), then the ICL fraction is the physically
meaningful parameter in comparison to cluster properties.

ICL fraction is apparently correlated with both mass and redshift
(Figure \ref{fig:massratio} \& \ref{fig:zratio}) and not with density
or total galaxy flux.  The selection
effect will again work against the predicted trend of ICL fraction to
increase with increasing mass \citep{murante2004,lin2004} and
increasing density.  Therefore the lack of trends of ICL fraction with
mass and density could be attributable to the selection bias.

\subsubsection{ICL fraction vs. Mass}
\label{mass2}

We find no trend in ICL fraction with mass.  Our data for ICL fraction
as a function of mass is inconsistent with the theoretical predictions
of \citet{murante2004}, \citet{murante07} (based on a cosmological
hydrodynamical simulation including radiative cooling, star formation,
and supernova feedback), and \citet{lin2004}(based on a model of
cluster mass and the luminosity of the BCG).  However
\citet{murante07} show a large scatter of ICL fractions within each
mass bin.  They also discuss the dependence of a simulations mass
resolution on the ICL fraction.  These theoretical predictions are
over-plotted on Figure \ref{fig:massratio}. Note that the simulations
generally report the fractional light in the ICL out to much larger
radii ($r_{virial}$ or $r_{200}$) than its surface brightness can be
measured observationally.  To compare the theoretical predictions at
$r_{virial}$ to our measurement at $0.25r_{virial}$, the predicted
values should be raised by some significant amount which depends on
the ICL and galaxy light profiles at large radii.  This makes the
predictions and the data even more inconsistent than it first appears.
As an example of the differences, a cluster with the measured ICL
fraction of A3888 would require a factor of greater than 100 lower
mass than the literature values to fall along the predicted trend.
Although these clusters are not dynamically relaxed, such large errors
in mass are not expected.  As an upper limit on the ICL flux, if we
assumed the entire cD galaxy was made of intracluster stars, that flux
plus the measured ICL flux would still not be enough to raise the ICL
fractions to the levels predicted by these authors.

There are no evident correlations between velocity dispersion and ICL
characteristics, although velocity dispersion is a mass estimator.
Large uncertainties are presumably responsible for the lack of
correlation.

%-------------------------------------
\subsubsection{ICL fraction vs. Redshift}

Figure \ref{fig:zratio} is a plot of redshift versus ICL fraction for
both the $r-$ and $B- $or $V-$bands.  We find a marginal
anti--correlation between ICL fraction and redshift with a very
shallow slope, if at all, in the direction that low redshift clusters
have higher ICL fractions (Spearman rank coefficient of -0.43). This
relation is strengthened when assuming fractions of blue galaxies are
higher in the higher redshift clusters(spearman rank of -0.6) (see \S
\ref{iclfraction}).  A trend of ICL fraction with redshift tells us about
the timescales of the mechanisms involved in stripping stars from
galaxies.  This relation is possibly affected by the same redshift
selection effects as discussed above.

Over the redshift range of our clusters, $0.31 > z > 0.05$, a
chi--squared fit to our data gives a range of fractional flux of 11 to
14\%.  \citet{willman2004} find the ICL fraction grows from 14 to
19\%.  over that same redshift range.  \citet{willman2004} measure the
ICL fraction at $r_{200}$ which means these values would need to be
increased in order to directly compare with our values.  While their
normalization of the relation is not consistent with our data, the
slopes are roughly consistent, with the caveat of the selection
effect.  The discrepancy is likely, at least in part, caused by
different definitions of ICL.  Simulations tag those particles which
become unbound from galaxies whereas in practice we do not have that
information and instead use surface brightness cutoffs and ICL profile
shapes.  \citet{rudick2006} do use a surface brightness cutoff in
their simulations to tag ICL stars which is very similar to our
measurement.  They find on average from their 3 simulated clusters a
change of ICL fraction of approximately 2\% over this redshift range.
We are not able to observationally measure such a small change in
fraction.  \citet{rudick2006} predict that in order to grow the ICL
fraction by 10\%, on average, we would need to track clusters as they
evolve from a redshift of 2 to the present.  However, both
\citet{willman2004} and \citet{rudick2006} find that the ICL fraction
makes small changes over short timescales (as major mergers or
collisions occur).

%-------------------------------------------------------------------
\subsection{ICL color}
The average color of the ICL, is roughly the same as the color of the
red ellipticals in each of the clusters. In \S8.1 of paper I we
discuss the implications of this on ICL formation redshift and
metallicity.  \citet{zibetti2005} have summed $g-$, $r-$, and $i-$
band imaging of 680 clusters in a redshift range of 0.2 - 0.3.
Similar to our results, they find that the summed ICL component has
roughly the same $g-r$ color at all radii as the summed cluster
population including the galaxies.  Since we have applied an
evolutionary correction to the ICL colors, if there is only passive
color evolution, the ICL will show no trend with redshift.  Indeed we
find no correlation between $B-r$ color and the redshift of the
cluster, as shown in Figure \ref{fig:zcolor} ($B-r = 2.3 \pm 0.2
(1\sigma)$).  ICL color may have the ability to broadly constrain the
epoch at which these stars were stripped.  In principle, as mentioned
in the introduction, we could learn at which epoch the ICL had been
stripped from the galaxies based on its color relative to the galaxies
assuming passively evolving ICL and ongoing star formation in
galaxies.  While this simple theory should be true, the color
difference between passively evolving stars and low star forming
galaxies may not be large enough to detect since clusters are not made
up of galaxies which were all formed at a single epoch and we don't
know the star formation rates of galaxies once they enter a cluster.

ICL color may have the ability to determine the types of galaxies from
which the stars are being stripped. Unfortunately the difference in
color between stars stripped from ellipticals, and for example stars
stripped from low surface brightness dwarfs is not large enough to
confirm in our data given the large amount of scatter in the color of
the ICL (see paper I for a more complete discussion).

There is no correlation in our sample between the presence or
direction of ICL color gradients and any cluster properties.  This is
very curious since we see both blue-ward and red-ward color gradients.
A larger sample with more accurate colors and without a selection bias
might be able to determine the origin of the color gradients.

%----------------------------------------------------------
\subsection{Profile Shape}
\label{profshape2}

\begin{figure}
%%\epsscale{1.0}
\includegraphics{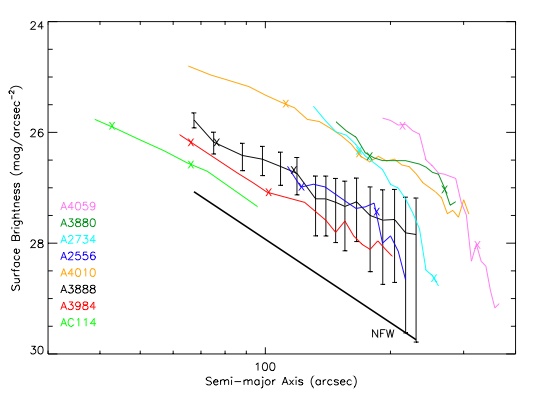}
\caption[Surface brightness profiles for the eight clusters with a
measurable profile]{Surface brightness profiles for the eight clusters
  with a measurable profile.  Profiles are listed on the plot in order
  of ascending redshift.  To avoid crowding, error bars are only
  plotted on one of the profiles.  Errors on the other profiles are
  similar at similar surface brightnesses.  All surface brightnesses
  have been shifted to $z = 0$ using surface brightness dimming, k,
  and evolutionary corrections. The x-axis remains in arcseconds and
  not in Mpc since the y-axis is in reference to arcseconds.  Physical
  scales are noted on the individual plots (\ref{fig:A4059} -
  \ref{fig:A118}).  In addition marks have been placed on each profile
  at the distances corresponding to 200kpc and 300kpc.  Also included
  as the solid black line near the bottom of the plot is a Hubble
  Reynolds surface brightness profile as a proxy for an NFW density
  profile with a scale length of 100kpc.  The ICL does not have a
  single uniform amount of flux or profile shape.  Profile shape does
  correlate with dynamical age where those clusters with steeper
  profiles are dynamically more relaxed (see \S \ref{profshape2}). }
\label{fig:predict}
%\epsscale{1}
\end{figure}

Figure \ref{fig:predict} shows all eight surface brightness profiles
for clusters that have central ICL components.  To facilitate
comparison, we have shifted all surface brightnesses to a redshift of
zero, including a correction for surface brightness dimming, a
k--correction, and an evolution correction.  We see a range in ICL
profile shape from cluster to cluster.  This is consistent with the
range of scale-lengths found in other surveys \citep[][find a range of
scale lengths from 18 - 480 kpc, fairly evenly distributed between 30
and 250 kpc]{gonzalez2005} .

\begin{figure}
%%\epsscale{.40}
\includegraphics[scale=0.60]{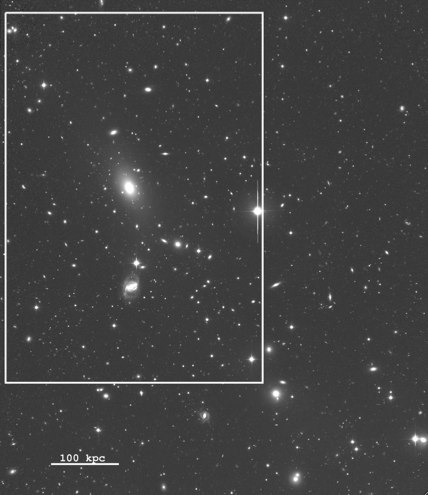}
\includegraphics[scale=0.78]{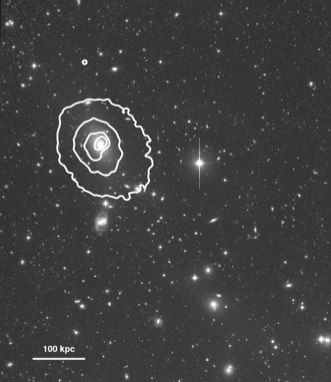}
\includegraphics[scale=1.2]{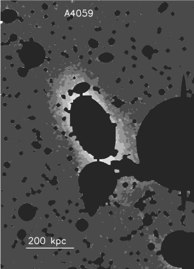}
\includegraphics[scale=0.70]{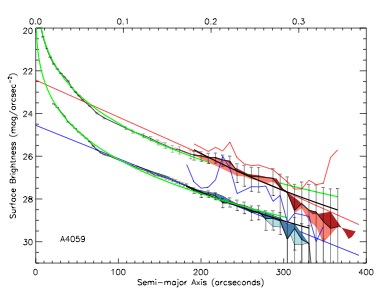}
\caption[A4059]{\footnotesize{A4059.  The plots moving left to right
    and top to bottom are as follows.  The first is our final combined
    $r-$band image zoomed in on the central cluster region.  The
    second plot shows X--ray isophotes where available.  Some clusters
    were observed during the ROSAT all sky survey, and so have X--ray
    luminosities, but have not had targeted observations to allow
    isophote fitting.  Isophote levels are derived from quick-look
    images taken from HEASARC.  X-ray luminosities of these clusters
    are listed in Table 1 of paper I and are discussed in the
    appendix.  The third plot shows our background subtracted, fully
    masked $r-$band image of the central region of the cluster,
    smoothed to aid in visual identification of the surface brightness
    levels.  Masks are shown in their intermediate levels which are
    listed in column 7 of Table \ref{tab:obspars2}.  The six
    gray-scale levels show surface brightness levels of up to 28.5,
    27.7,27.2,26.7 \magarc.  The fourth plot shows the surface
    brightness profiles of the ICL (surrounded by shading;$r-$band on
    top, $V-$ or $B-$band on the bottom) and cluster galaxies as a
    function of semi-major axis.  The bottom axis is in arcseconds and
    the top axis corresponds to physical scale in Mpc.  Error bars
    represent the $1\sigma$ background identification errors as
    discussed in \S \ref{noise2}.  DeVaucouleurs fits to the entire cD
    plus ICL profile are over-plotted.  }}
\label{fig:A4059}
%\epsscale{1}
\end{figure}

\begin{figure}
%%\epsscale{.45}
\includegraphics[scale=0.6]{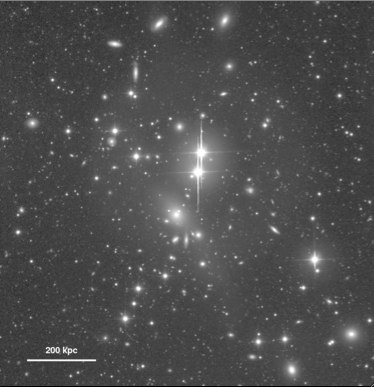}
%\includegraphics{f4b.jpg}
%\epsscale{0.42}
\includegraphics[scale=0.9]{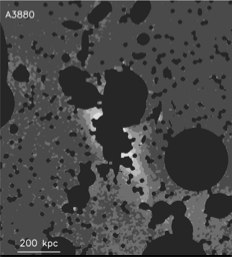}
%\epsscale{0.6}
\centering
\includegraphics[scale=0.70]{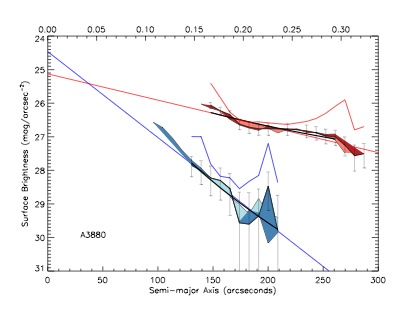}
\caption[A3880]{A3880, same as Figure \ref{fig:A4059}}
\label{fig:A3880}
%\epsscale{1}
\end{figure}

\begin{figure}
%\epsscale{.45}
\includegraphics[scale=0.8]{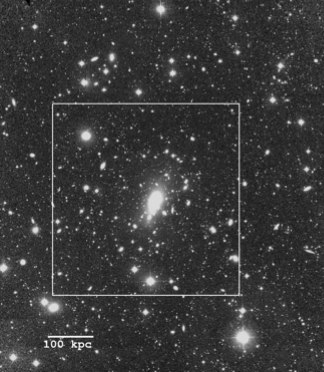}
\includegraphics[scale=0.8]{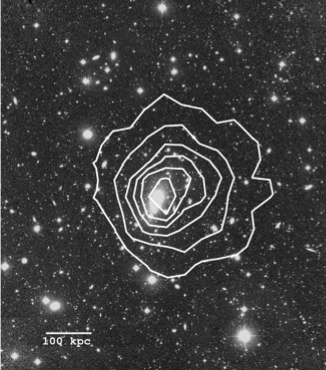}
\includegraphics{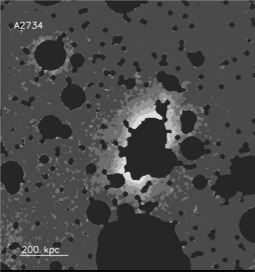}
\includegraphics[scale=0.70]{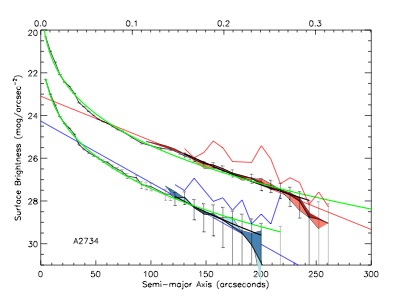}
\caption[A2734]{A2734, same as Figure \ref{fig:A4059}}
\label{fig:A2734}
%\epsscale{1}
\end{figure}

\begin{figure}
%\epsscale{.55}
\centering
\includegraphics[scale=0.48]{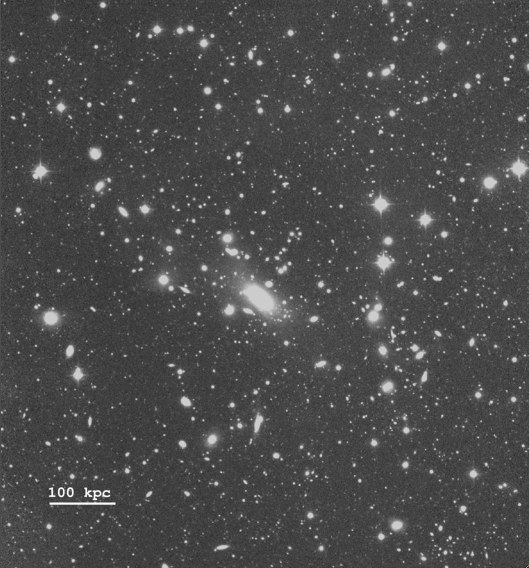}
%\includegraphics{f6b.jpg}
%\epsscale{0.44}
\includegraphics[scale=0.71]{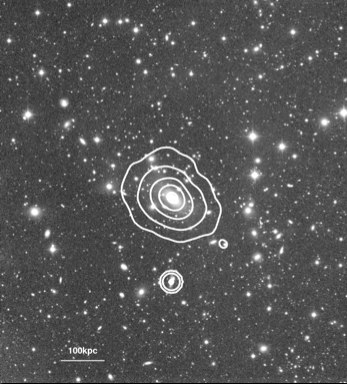}
%\epsscale{0.45}
\includegraphics{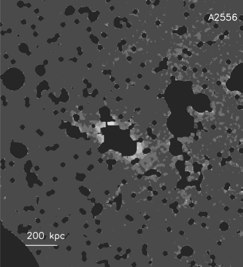}
\includegraphics[scale=0.65]{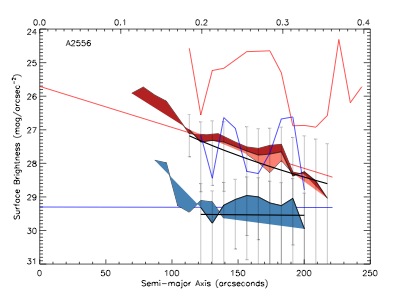}
\caption[A2556]{A2556, same as Figure \ref{fig:A4059}}
\label{fig:A2556}
%\epsscale{1}
\end{figure}

\begin{figure}
%\epsscale{.45}

\includegraphics[scale=0.68]{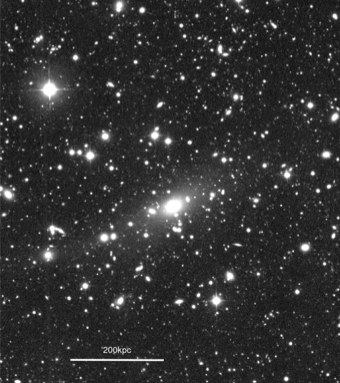}
%\includegraphics{f7b.jpg}
%\epsscale{0.5}
\includegraphics[scale=0.96]{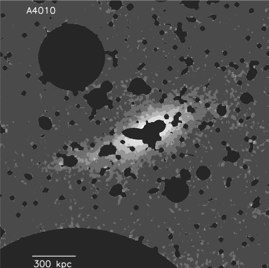}
\centering
\includegraphics{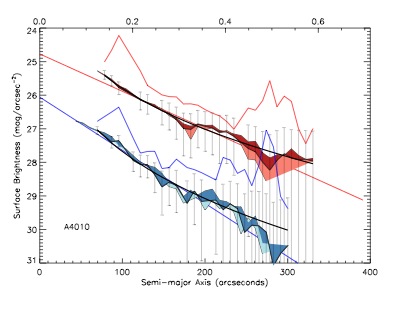}
\caption[A4010]{A4010, same as Figure \ref{fig:A4059}}
\label{fig:A4010}
%\epsscale{1}
\end{figure}

\begin{figure}
%\epsscale{.45}
\includegraphics[scale=0.5]{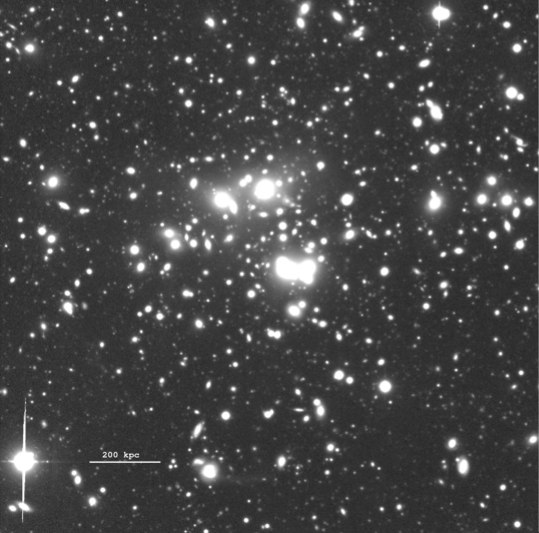}
\includegraphics[scale=0.5]{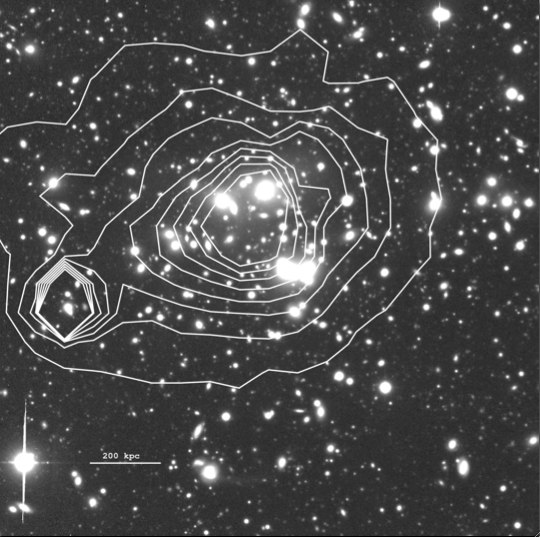}
\includegraphics[scale=1.0]{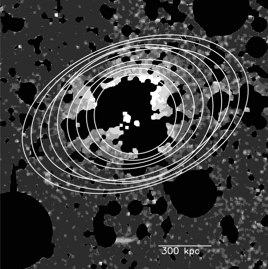}
\includegraphics[scale=0.65]{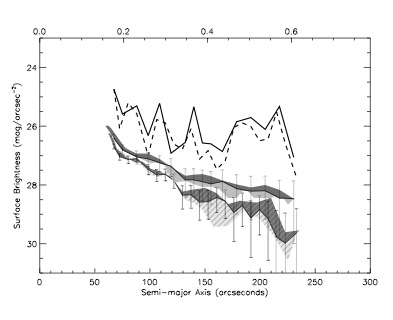}
\caption[A3888]{A3888, same as Figure \ref{fig:A4059}, except here we
  show the elliptical isophotes of the ICL over-plotted on the surface
  brightness image.}
\label{fig:A3888}
%\epsscale{1}
\end{figure}

\begin{figure}
%\epsscale{.45}
\centering
\includegraphics[scale=0.81]{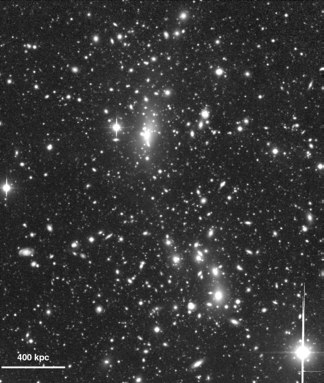}
%\includegraphics{f9b.jpg}
%\epsscale{0.405}
\includegraphics[scale=1.15]{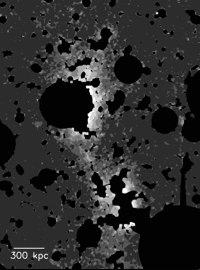}
%\epsscale{0.6}
\includegraphics{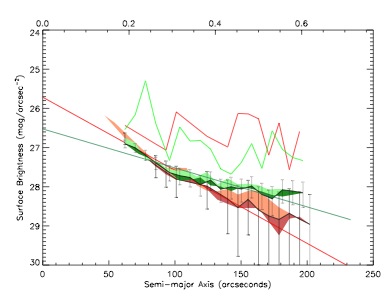}
\caption[A3984]{A3984, same as Figure \ref{fig:A4059}}
\label{fig:A3984}
%\epsscale{1}
\end{figure}

\begin{figure}
%\epsscale{.4}
\centering
\includegraphics[scale=0.6]{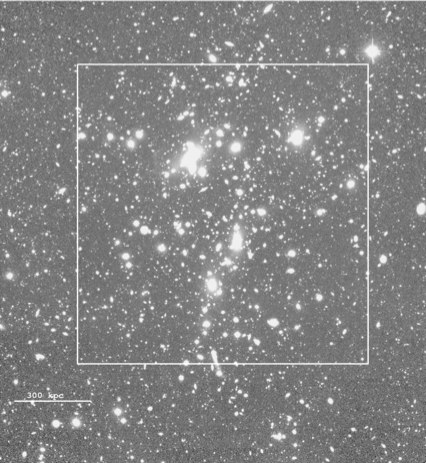}
\includegraphics[scale=0.6]{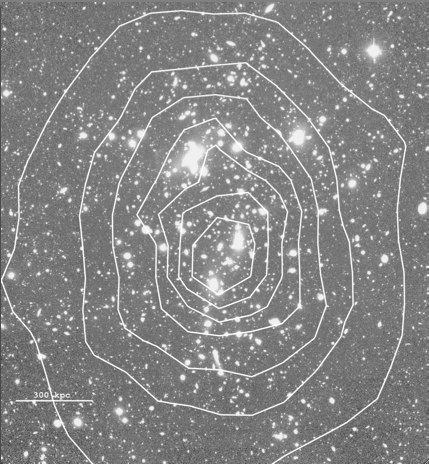}
\includegraphics{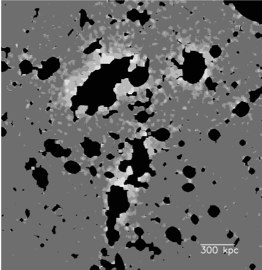}
\caption[A141]{A141, same as Figure \ref{fig:A4059}, except we are not
able to measure a surface brightness profile or consequently a color
profile.}
\label{fig:A141}
%\epsscale{1}
\end{figure}

\begin{figure}
%\epsscale{.45}
\centering
\includegraphics[scale=0.5]{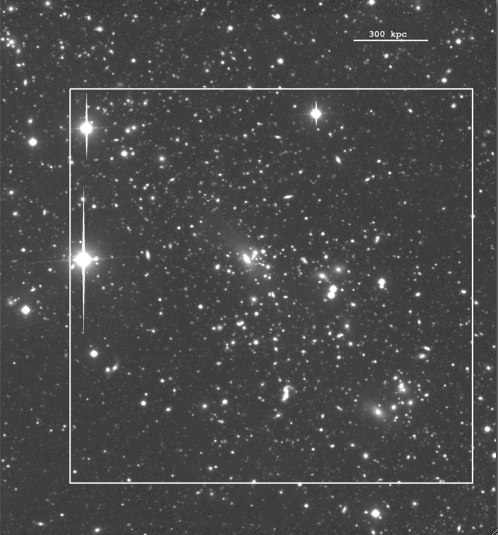}
\includegraphics[scale=0.5]{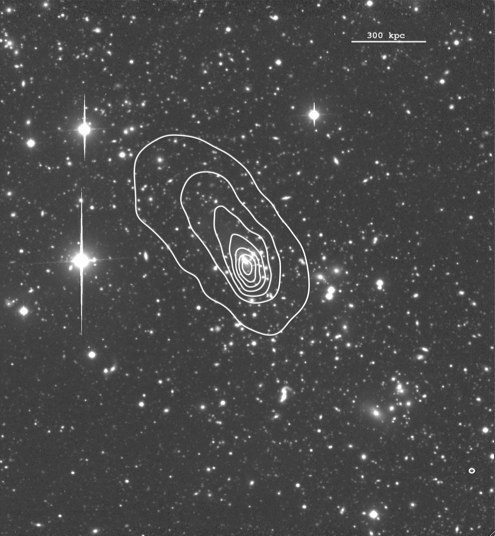}
\includegraphics[scale=0.99]{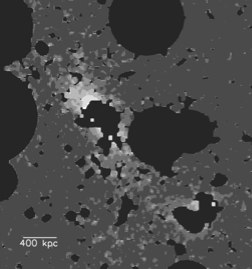}
\includegraphics[scale=0.62]{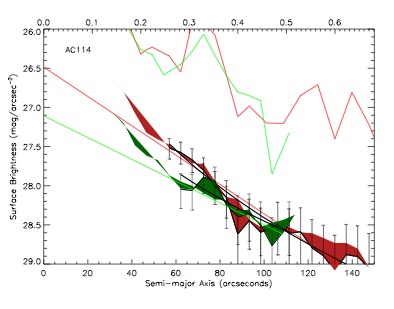}
\caption[A114]{A114, same as Figure \ref{fig:A4059}}
\label{fig:A114}
%\epsscale{1}
\end{figure}

\begin{figure}
%\epsscale{.45}
\centering
\includegraphics[scale=0.5]{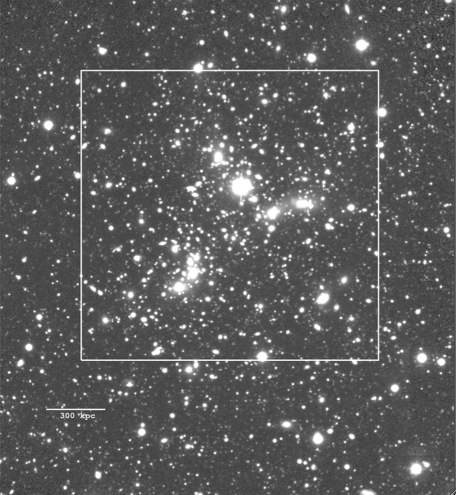}
\includegraphics[scale=0.5]{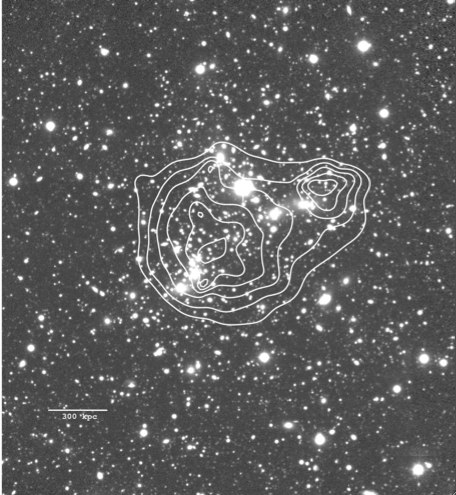}
\includegraphics[scale=0.9]{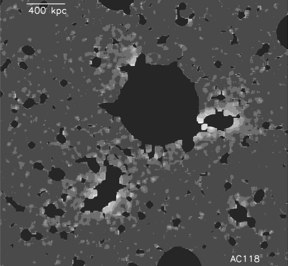}
\caption[A118]{A118, same as Figure \ref{fig:A4059}, except we are not
able to measure a surface brightness profile or consequently a color
profile.}
\label{fig:A118}
%\epsscale{1}
\end{figure}

\begin{figure}
%\epsscale{.4}
\centering
\includegraphics[scale=0.5]{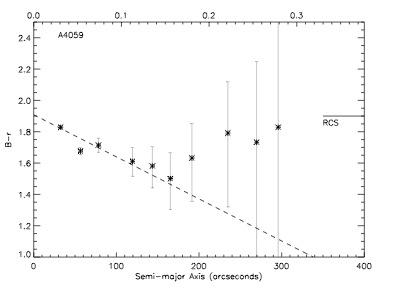}
\includegraphics[scale=0.5]{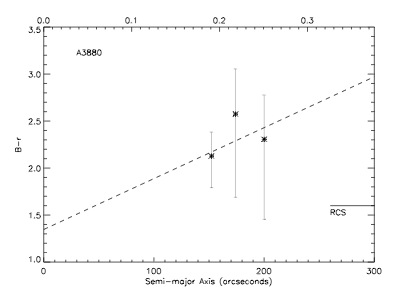}
\includegraphics[scale=0.5]{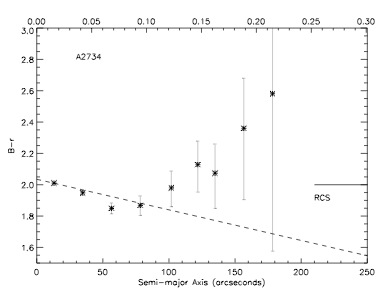}
\includegraphics[scale=0.5]{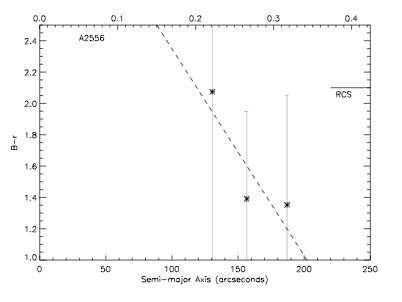}
\includegraphics[scale=0.5]{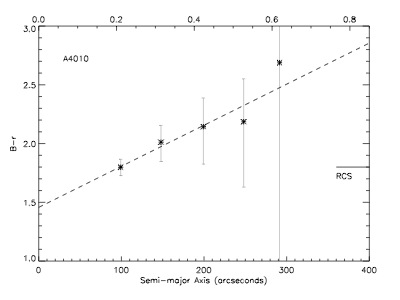}
\includegraphics[scale=0.5]{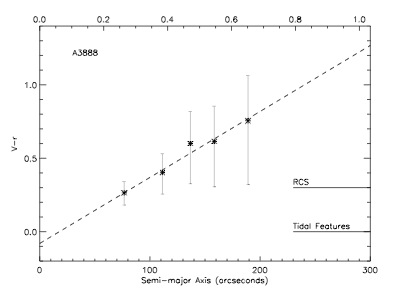}
\includegraphics[scale=0.5]{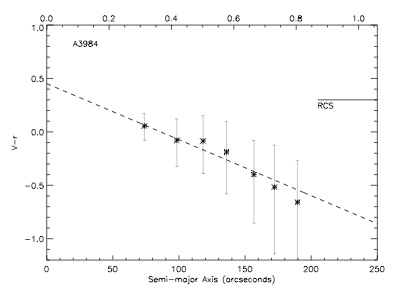}
\includegraphics[scale=0.5]{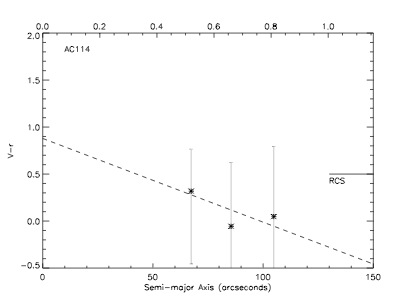}
\caption[allcolor]{The color profile of the eight clusters where
  measurement was possible plotted as a function of semi-major axis in
  arcseconds on the bottom and Mpc on the top. The average color of
  the red cluster sequence is shown for comparison, as well as the
  best fit linear function to the data.}
\label{fig:allcolor}
%\epsscale{1}
\end{figure}

\begin{figure}
\centering
%\epsscale{0.8}
\includegraphics{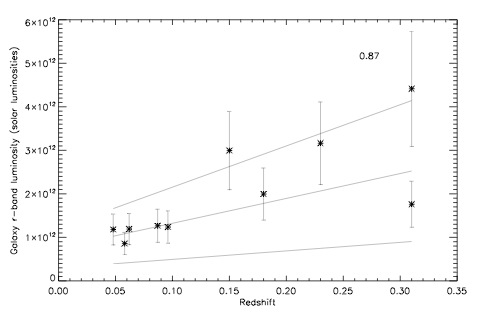}
\caption[Redshift versus total galaxy flux within one quarter of a
  virial radius]{Redshift versus total galaxy flux within one quarter
  of a virial radius. The Spearman rank coefficient is printed in the
  upper right corner.  The best fit linear function as well as the
  lines representing $\pm 2 \sigma$ in both slope and y intercept are
  also plotted.  The strong correlation between redshift and total
  galaxy flux shows the incompleteness of the Abell sample which does
  not include high-redshift, low-flux clusters}
\label{fig:clusterz}
%\epsscale{1}
\end{figure}

\begin{figure}
\centering
%\epsscale{0.8}
\includegraphics{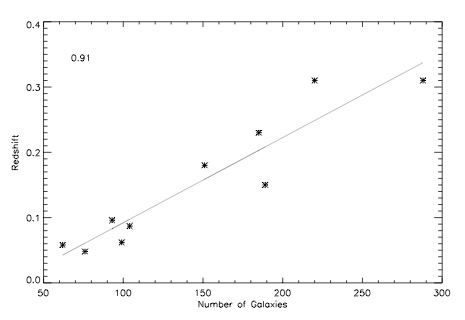}
\caption[Projected number of galaxies versus redshift]{Projected
number of galaxies versus redshift.  Galaxies brighter than $M_r =
-18.5$ within 800 \h kpc are included in this count, which is used as
a proxy for density.  The Spearman rank coefficient is printed in the
upper left corner.  There is a strong correlation between density and
redshift.  The best fit linear function is included.  While we do
expect clusters to become less dense over time, this strong
correlation is not expected.  Instead this is due to an incompleteness
at high redshift.  See \S \ref{discuss2} for a discussion of the
effects of this selection effect.}
\label{fig:ngalsz}
%\epsscale{1}
\end{figure}

\begin{figure}
\centering
%\epsscale{0.8}
\includegraphics{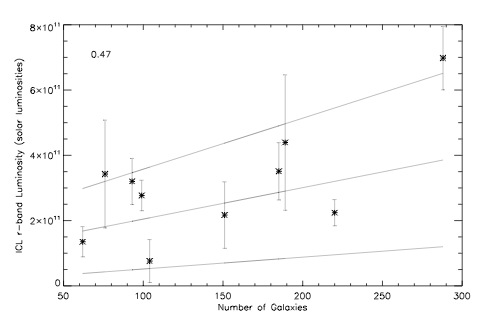}
\caption[Projected number of galaxies versus ICL luminosity]{Projected
number of galaxies versus ICL luminosity.  ICL luminosity shows
$1\sigma$ error bars and has been K and evolution corrected.  Galaxies
brighter than $M_r = -18.5$ within 800 \h kpc are included in this
count, which is used as a proxy for density.  The Spearman rank
coefficient is printed in the upper left corner. The best fit linear
function as well as the lines representing $\pm 2 \sigma$ in both
slope and y intercept are also plotted. There is a mild correlation
between density and ICL luminosity such that higher density clusters
have a larger amount of ICL flux.  }
\label{fig:ngalsicl}
%\epsscale{1}
\end{figure}

\begin{figure}
\centering
%\epsscale{0.8}
\includegraphics{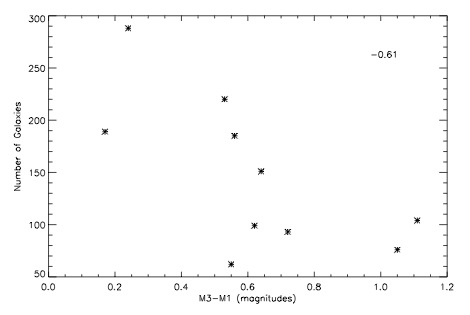}
\caption[The difference in magnitude between the first and third
ranked galaxy versus projected number of galaxies brighter than $M_r =
-18.5$ within 800 \h kpc]{The difference in magnitude between the
first and third ranked galaxy versus projected number of galaxies
brighter than $M_r = -18.5$ within 800 \h kpc, which is used as a
proxy for density. Clusters with cD galaxies will have larger M3-M1
values.  This plot implies that over time galaxies merge in clusters
to make a cD galaxy, and by the time the cD galaxy has formed, the
global density is lower.  As discussed in the \S \ref{discuss2}, we
assume this is not a selection bias.}
\label{fig:ngalsm3m1}
%\epsscale{1}
\end{figure}

\begin{figure}
\centering
%\epsscale{0.8}
\includegraphics{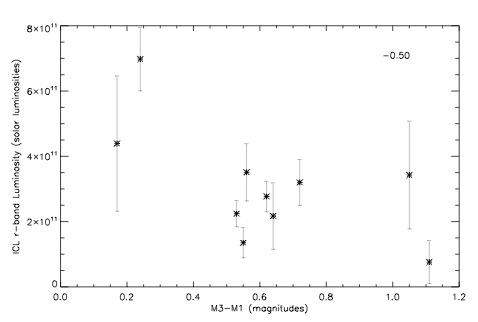}
\caption[The difference in magnitude between the first and third
ranked galaxy versus ICL luminosity]{The difference in magnitude
between the first and third ranked galaxy versus ICL luminosity. ICL
luminosity shows $1\sigma$ error bars and has been K and evolution
corrected.  Clusters which have cD galaxies have larger M3 - M1 values
and are dynamically older clusters.  There is a mild correlation
between dynamic age and ICL luminosity indicating that the ICL evolves
at roughly the same rate as the cluster.}
\label{fig:m3m1icl}
%\epsscale{1}
\end{figure}

\begin{figure}
\centering
%\epsscale{0.8}
\includegraphics{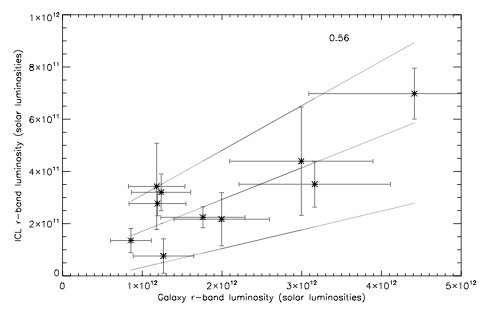}
\caption[The flux in galaxies versus the flux in ICL in units of solar
  luminosities]{The flux in galaxies versus the flux in ICL in units
  of solar luminosities.  Errors on ICL luminosity are $1\sigma$.
  Errors on galaxy luminosity are 30\% as estimated in \S
  \ref{member2}.  Over-plotted is the best fit linear function as well
  as two lines which represent $2\sigma$ errors in both y-intercept
  and slope.  The Spearman rank coefficient is printed in the upper
  right.  Here galaxy luminosity is assumed to be a proxy for mass, so
  we find a significant correlation between mass and ICL flux such
  that more massive clusters have a larger amount of ICL flux.}
\label{fig:clustericl}
%\epsscale{1}
\end{figure}

\begin{figure}
\centering
%\epsscale{0.8}
\includegraphics{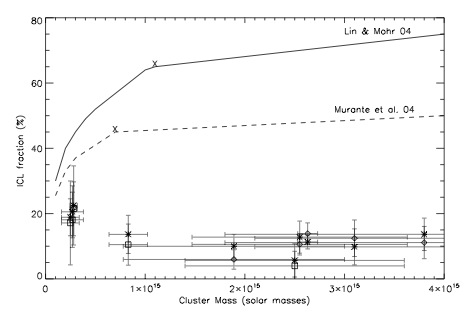}
\caption[Cluster mass versus the ICL fraction measured at one quarter
of the virial radius]{Cluster mass versus the ICL fraction measured at
  one quarter of the virial radius.  Stars denote the $r-$band while
  squares show $B-$ and diamonds show $V-$band.  Errors on ICL
  fraction are $1\sigma$ as discussed in \S \ref{noise2}.  Mass
  estimates and errors are taken from the literature as discussed in
  \S \ref{A4059} - \S \ref{AC118}.  The predictions of \citet{lin2004}
  and \citet{murante2004} at the virial radius are shown for
  comparison.  These represent extrapolations beyond roughly $1\times
  10^{15}$ \msun\ in both cases (as marked by the crosses).  The
  roughly constant ICL fraction with mass can be explained using
  hierarchical formation by the in-fall of groups with a similar ICL
  fraction as the main cluster, or by increased interaction rates with
  the infall of the groups, or both.}
\label{fig:massratio}
%\epsscale{1}
\end{figure}

\begin{figure}
\centering
%\epsscale{0.8}
\includegraphics{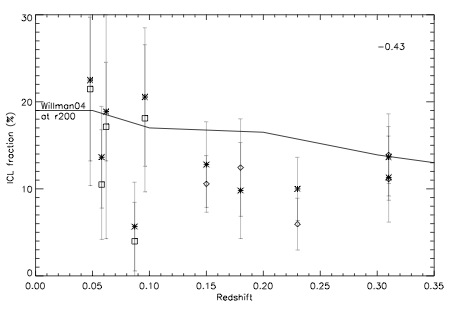}
\caption[Cluster redshift versus ICL fraction measured at one quarter
  of the virial radius]{Cluster redshift versus ICL fraction measured
  at one quarter of the virial radius. As in Figure
  \ref{fig:massratio}, starred symbols denote the $r-$band, squares
  show $B-$band, and diamonds show $V-$band fractions. The prediction of
  \citet{willman2004} for the ICL fraction as measured at $r_{200}$ is
  shown for comparison. This prediction would increase if measured at
  smaller radii, such as was used in our measurement.  There is mild
  evidence for a correlation between redshift and ICL fraction such
  that ICL fraction grows with decreasing redshift.  This trend is
  consistent with ongoing ICL formation.}
\label{fig:zratio}
%\epsscale{1}
\end{figure}

\begin{figure}
\centering
%\epsscale{0.8}
\includegraphics{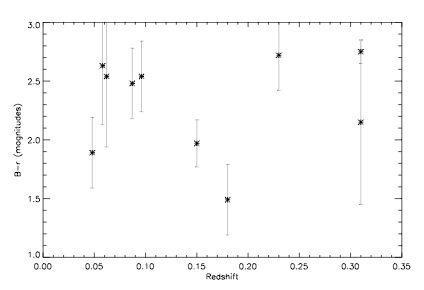}
\caption[Cluster redshift versus ICL color in $B-r$ which has been k
  corrected and had simple passive evolution applied to it]{Cluster
  redshift versus ICL color in $B-r$ which has been k corrected and
  had simple passive evolution applied to it. If a color gradient is
  detected in a given cluster then the mean color plotted here is that
  measured near the center of the profile, weighted slightly toward
  the center. There is no trend in redshift with ICL color which leads
  to the conclusion that the ICL is simply passively reddening. }
\label{fig:zcolor}
%\epsscale{1}
\end{figure}

The profiles are equally well fit with the empirically motivated
deVaucouleurs profiles and simple exponential profiles which are shown
in the individual profile plots in Figures \ref{fig:A4059} -
\ref{fig:A118}.  The profiles can also be fit with a Hubble--Reynolds
profile which is a good substitute for the more complicated surface
brightness profile of an NFW density profile \citep{lokas2001}.  An
example of this profile shape is shown in Figure \ref{fig:predict}
with a 100 kpc scale length defined as the radius inside of which the
profile contains 25\% of the luminosity.  This profile shape is what
you would predict given a simple spherical collapse model.  The
physically motivated Hubble--Reynolds profile gives acceptable fits to
the ICL profiles with the exception of A4059, A2734, \& A2556 which
have steeper profiles.  We explore causes of the differing profile
shapes for these three clusters.

A steeper profile is correlated with M3-M1, density, total cluster
flux, and redshift.  These three clusters have an average M3-M1 value
of $0.93 \pm 0.27$ as compared to the average of $0.49 \pm 0.20$ for
the remaining 7 clusters.  These three clusters are also three of the
four lowest redshift clusters, have an average of 93 galaxies which is
45\% smaller than the value for the remaining sample, and have an
average cluster flux of $12.3\times10^{11}$\lsun which is 47\% smaller
than the value for the remaining sample.

We have the same difficulties here in distinguishing between the
selection effects and the true physical correlations.  The key
difference is that the three clusters with the steepest profiles are
the most relaxed clusters (which is not a redshift selection effect).
We use ``most relaxed'' to describe the three clusters with the most
symmetric X--ray isophotes that have single, central, smooth ICL
profiles.  This is consistent with our finding that M3-M1 is a key
indicator of ICL flux in \S\ref{m3m1} and that ICL can form either in
groups at early times or at later times through galaxy interactions in
the dense part of the cluster.  If galaxy groups in which the ICL
formed are able to get to the cluster center then their ICL will also
be found in the cluster center, and can be hiding in the cD galaxy.
If the galaxy groups in which the ICL formed have not coalesced in the
center then the ICL will be less centrally distributed and therefore
have a shallower profile.  This is consistent with the recent
numerical work by \citet{murante07} who find that the majority of the
ICL is formed by the merging processes which create the BCG's in
clusters.  This process leads o the ICL having a steeper profile shape
than the galaxies and having greater than half of the ICL be located
inside of $250$\h kpc, approaching radii where we do not measure the
ICL due to the presence of the BCG.  Their simulations also confirm
that different clusters with different dynamical histories will have
differing amounts and locations of ICL.

%-------------------------------------------------------------------------

%centers

\section{Conclusion}
\label{conclude2}

We have identified an intracluster light component in all 10 clusters
which has fluxes ranging from $0.76\times 10^{11}$ to $7.0\times
10^{11}$ \h \lsun in $r$ and $0.14\times 10^{11}$ to $1.2\times
10^{11}$ \h \lsun in the $B-$band, ICL fractions of 6 to 22\% of the
total cluster light within one quarter of the virial radius in $r$
and 4 to 21\% in the $B-$band, and $B-r$ colors ranging from 1.49 to
2.75 magnitudes.  This work shows that there is detectable ICL in
clusters and groups out to redshifts of at least 0.3, and in two bands
including the shorter wavelength $B-$ or $V-$band.

The interpretation of our results is complicated by small number
statistics, redshift selection effects of Abell clusters, and the fact
that the ICL is evolving with time.  Of the cluster properties (M3-M1,
density, redshift, and cluster flux), only M3-M1 and redshift are not
correlated.  As a result of these selection effects ICL flux is
apparently correlated with density and total galaxy flux but not with
redshift or mass and ICL fraction is apparently correlated with
redshift but not with M3-M1, density, total galaxy flux, or mass.
However, we do draw conclusions from the ICL color, average values of
the ICL fractions, the relation between ICL flux and M3-M1, and the
ICL profile shape.

We find a passively evolving ICL color which is similar to the color
of the RCS at the redshift of each cluster. The relations between ICL
fraction with redshift and ICL fraction with mass show the
disagreement of our data with simulations since our fractional fluxes
are lower than those predictions.  These discrepancies do not seem to
be caused by the details of our measurement.

Furthermore we find evidence that clusters with symmetric X--ray
profiles and cD galaxies have both less ICL flux and significantly
steeper profiles.  The lower amount of flux can be explained if ICL
stars have become indistinguishable from cD stars.  As the cluster
formed a cD galaxy any groups which participated in the merging
brought their ICL stars with them, as well as created more ICL through
interactions.  If a cD does not form, then the ICL already in groups
or actively forming is also prevented from becoming very centralized
as it has no way of loosing energy or angular momentum on its own.
While the galaxies or groups are subject to tidal forces and dynamical
friction, the ICL, once stripped, will not be able to loose energy
and/or angular momentum to these forces, and instead will stay on the
orbit on which it formed.

Observed density may not be a good predictor of ICL properties since
it does not directly indicate the density at the time in which the ICL
was formed.  We do indeed expect density at any one epoch to be linked
to ICL production at that epoch through the interaction rates.

The picture that is emerging from this work is that ICL is ubiquitous,
not only in cD clusters, but in all clusters, and in group
environments.  The amount of light in the ICL is dependent upon
cluster morphology.  ICL forms from ongoing processes including
galaxy--galaxy interactions and tidal interactions with the cluster
potential \citep{moore1996,gnedin2003a} as well as in groups
\citep{rudick2006}.  With time, as multiple interactions and
dissipation of angular momentum and energy lead groups already
containing ICL to the center of the cluster, the ICL moves with the
galaxies to the center and becomes indistinguishable from the cD's
stellar population.  Any ICL forming from galaxy interactions stays on
the orbit where it was formed.

A large, complete sample of clusters, including a proportionate amount
with high redshift and low density, will be able to break the
degeneracies present in this work.  Shifting to a lower redshift range
will not be as beneficial because a shorter range than presented here
will not be large enough to see the predicted evolution in the ICL
fraction.

In addition to large numbers of clusters it would be beneficial to go
to extremely low surface brightness levels ($< \sim 30$ \magarc) to reduce
significantly the error bars on the color measurement and thereby
learn about the progenitor galaxies of the ICL and the timescales for
stripping.  It will not be easy to achieve these surface brightness
limits for a large sample which includes high-redshift low-density
clusters since those clusters will have very dim ICL due to both an
expected lower amount as correlated with density, and due to surface
brightness dimming.
%-------------------------------------------------------------------------------

\acknowledgments

We acknowledge J. Dalcanton and V. Desai for observing support and
R. Dupke, E. De Filippis, and J. Kempner for help with X--ray data.  We
thank the anonymous referee for useful suggestions on the manuscript.
Partial support for J.E.K. was provided by the National Science
Foundation (NSF) through UM's NSF ADVANCE program.  Partial support
for R.A.B. was provided by a NASA Hubble Fellowship grant
HF-01088.01-97A awarded by Space Telescope Science Institute, which is
operated by the Association of Universities for Research in Astronomy,
Inc., for NASA under contract NAS 5-2655.  This research has made use
of data from the following sources: USNOFS Image and Catalogue Archive
operated by the United States Naval Observatory, Flagstaff Station
(http://www.nofs.navy.mil/data/fchpix/); NASA/IPAC Extragalactic
Database (NED), which is operated by the Jet Propulsion Laboratory,
California Institute of Technology, under contract with the National
Aeronautics and Space Administration; the Two Micron All Sky Survey,
which is a joint project of the University of Massachusetts and the
Infrared Processing and Analysis Center/California Institute of
Technology, funded by the National Aeronautics and Space
Administration and the National Science Foundation; the SIMBAD
database, operated at CDS, Strasbourg, France; and the High Energy
Astrophysics Science Archive Research Center Online Service, provided
by the NASA/Goddard Space Flight Center.

%------------------------------------------------------------------------
\bibliography{ms.bbl}  
%\bibliography{jkrick}  

%============================================
\clearpage

%\begin{sidewaystable*}[htbp]
%\scriptsize
%\begin{center}

%\begin{tabular}{l c c c c c c c c c c c c }
%\hfil & \hfil  & \hfil & \hfil & \hfil & \hfil & \hfil & \hfil \\
%\tableline
%\tableline
%\\

\begin{deluxetable}{l c c c c c c c c c c c c c c}
\tabletypesize{\tiny}
\rotate
\tablewidth{0pc}
\tablecolumns{15}
\tablecaption{Cluster characteristics\label{tab:characteristics2}}

%Column heading definitions:

\tablehead{
\colhead{Cluster} & 
\colhead{$z$} 		 & 
%\colhead{B/M} 	  &
\colhead{$M_3-M_1$}  &
\colhead{Richness} 	& 
\colhead{ngals} 	& 
\colhead{$\sigma_v$} 	& 
\colhead{$r_{virial}$} &
\colhead{Mass} &
\multicolumn{2}{c}{Cluster Flux} &
\multicolumn{2}{c}{ICL Flux} &
\multicolumn{2}{c}{Ratio} &
\colhead{ICL} 
\\
\colhead{name}    & 
\colhead{ }     &
%\colhead{ } 	  &
\colhead{mag} 	  &
\colhead{Class}	& 
\colhead{ }     &
\colhead{km/s} 	    & 
\colhead{Mpc}	& 
\colhead{$10^{14}$ \msun} &	
\colhead{B $10^{11}$ \lsun} &
\colhead{r $10^{11}$ \lsun} &
\colhead{B $10^{11}$ \lsun} &
\colhead{r $10^{11}$ \lsun} &
\colhead{B\%} &
\colhead{r\%} &
\colhead{color} 
}

%\\
%\tableline

\startdata
\hfil 	& \hfil    & \hfil    & \hfil	 & \hfil    & \hfil   & \hfil  & \hfil	 & \hfil   & \hfil    & \hfil 	\\
A4059$^{\dag}$ & 0.048$^{\rm r}$  & $1.05\pm.05$  & 1 & 76 & 845$^{\rm +280}_{\rm -140}$$^{\rm t}$ & 2.6$^{\rm i}$  & 2.82 $^{\rm +0.37}_{\rm -0.34}$$^{\rm i}$  & $4.2\pm1.3$ & $12\pm3.5$ & $1.2\pm.24$ & $3.4\pm1.7$ & $21\pm8$ & $22\pm12$ & 1.89\\
%\hfil 	       & \hfil          & \hfil & \hfil	& \hfil & \hfil  		 & \hfil  & \hfil	\\
A3880$^{\dag}$ & 0.058$^{\rm r}$  & $0.55\pm.05$  & 0 & 62 & 827$^{\rm +120}_{\rm -79}$$^{\rm m}$  & 2.5$^{\rm e}$ & 8.3 $^{\rm +2.8}_{\rm -2.1}$$^{\rm f}$  & $3.8\pm1.1$ & $8.6\pm2.6$ & $0.44\pm0.23$ & $1.4\pm0.46$ & $10\pm6$ & $14\pm6$ & 2.63\\
%\hfil 	       & \hfil          & \hfil & \hfil	& \hfil & \hfil  		 & \hfil  & \hfil	\\
A2734$^{\dag}$ 	& 0.062$^{\rm r}$  & $0.62\pm.05$ & 1 & 99 & 628$^{\rm +61}_{\rm -57}$$^{\rm m}$   & 2.4$^{\rm i}$ & 2.49 $^{\rm +0.89}_{\rm -0.63}$$^{\rm i}$  & $3.4\pm1.0$ & $12\pm3.6$ & $0.7\pm0.47$ & $2.8\pm0.47$ & $17\pm13$ & $19\pm6$ & 2.54\\
%\hfil 	       & \hfil          & \hfil & \hfil	& \hfil & \hfil  		 & \hfil  & \hfil	\\
A2556$^{\dag}$	& 0.087$^{\rm r}$  & $1.11\pm.05$ & 1 & 104 & 1247$^{\rm +249}_{\rm -249}$$^{\rm t}$ & 2.6$^{\rm e}$ & $25\pm1$$^{\rm g}$ & $3.3\pm0.99$ & $13\pm3.8$ & $0.14\pm0.14$ & $0.76\pm0.66$ & $4\pm4$ & $6\pm5$ & 2.48\\
%\hfil 	       & \hfil          & \hfil & \hfil	& \hfil & \hfil  		 & \hfil  & \hfil	\\
A4010$^{\dag}$	& 0.096$^{\rm r}$  & $0.72\pm.05$	& 1 & 93 & 625$^{\rm +127}_{\rm -95}$$^{\rm m}$  & 3.1$^{\rm e}$ & 3.8 $^{\rm +1.6}_{\rm -1.2}$$^{\rm f}$ & $3.5\pm1.0$ & $12\pm3.7$ & $0.77\pm0.28$ & $3.2\pm0.70$ & $18\pm8$ & $21\pm8$ & 2.54\\
%\hfil 	       & \hfil          & \hfil & \hfil	& \hfil & \hfil  		 & \hfil  & \hfil	\\
A3888	& 0.151$^{\rm r}$         & $0.17\pm.04$ & 2 & 189 & 1102$^{\rm +137}_{\rm -107}$$^{\rm n}$ & 3.7$^{\rm i}$ & 25.5 $^{\rm +10.5}_{\rm -7.4}$$^{\rm i}$ & $7.2\pm2.2$ & $30\pm9.0$ & $0.86\pm.25$ & $4.4\pm2.1$ & $11\pm3$ & $13\pm5$ & 1.97 \\
%\hfil 	       & \hfil          & \hfil & \hfil	& \hfil & \hfil  		 & \hfil  & \hfil	\\
A3984$^{\dag}$	& 0.181$^{\rm r}$  & $0.64\pm.04$ & 2 & 151	& \nodata      			   & 3.5$^{\rm e}$ & $31\pm10$$^{\rm c}$   & $4.4\pm1.3$ & $20\pm6.0$ & $0.62\pm0.21$ & $2.2\pm1.0$ & $12\pm6$ & $10\pm6$ & 1.49\\
%\hfil 	       & \hfil          & \hfil & \hfil	& \hfil & \hfil  		 & \hfil  & \hfil	\\
A0141$^{\dag}$	& 0.23$^{\rm r}$   & $0.56\pm.04$	& 3 & 185 & \nodata      			  & 3.7$^{\rm e}$ & 18.9 $^{\rm +11.1}_{\rm -8.7}$$^{\rm d}$ & $5.4\pm1.6$ & $32\pm9.5$ & $0.34\pm0.11$ & $3.5\pm0.88$ & $6\pm3$ & $10\pm4$ & 2.72\\
%\hfil 	       & \hfil          & \hfil & \hfil	& \hfil & \hfil  		 & \hfil  & \hfil	\\
AC114 & 0.31$^{\rm a}$ & $0.45\pm.04$ &2 & 220 & 1388$^{\rm +128}_{\rm-71}$$^{\rm n}$ & 3.5$^{\rm b}$ & 26.3$^{\rm+8.2}_{\rm-7.1}$$^{\rm n}$   & $2.3\pm0.70$ & $18\pm5.3$ & $0.38\pm0.08$ & $2.2\pm0.4$ & $14\pm3$ & $11\pm2$ & 2.15\\
%\hfil 	       & \hfil          & \hfil & \hfil	& \hfil & \hfil  		 & \hfil  & \hfil	\\
AC118 	& 0.308$^{\rm r}$         & $0.24\pm.04$   & 3 & 288 & 1947$^{\rm +292}_{\rm -201}$$^{\rm h}$ & 3.4$^{\rm b}$ & $38\pm37$$^{\rm n}$   & $5.4\pm1.6$ & $44\pm1.3$ & $0.67\pm0.17$ &$ 7.0\pm0.97$ & $11\pm5$ & $14\pm5$ & 2.75\\
\enddata
%\\
%\tableline
%\\
%\end{tabular}

{\footnotesize \addtolength{\baselineskip}{-5pt} {\bf Notes:} Sources
for the virial radii and mass are discussed in
\S\ref{A4059}-\ref{AC118} and generally come from X--ray data.  $\dag$:
We have obtained additional photometric and spectroscopic data for
these cluster to be published in a forthcoming paper) 
a: \citet{abell1989}.
b: \citet{allen1998}.
c: \citet{cypriano2004}.
d: \citet{dahle2002}.
e: \citet{ebeling1996}.
f: \citet{girardi1998a}.
g:\citet{reimers1996}.
%b: \citet{batuski1999}.
%c: \citet{busarello2002} 
%d: \citet{chen1998}.
%e: \citet{ciardullo1985}.
%f: \citet{collins1995}.
%g: \citet{couch1984}.
h: \citet{couch1987}.
i: \citet{reiprich2002}.
%i: \citet{couch2001}.
%j: \citet{denHartog1995}.
%k: \citet{DePropris2002}.
%l: \citet{ebeling1996} 
m: \citet{Girardi1998b}.
n: \citet{girardi2001}.
%o: \citet{Kowalski1983}.
%p: \citet{Mazure1996}.
%q: \citet{Stein1996}.
r: \citet{struble1999}.
%s: \citet{teague1990}.
t: \citet{Wu1999}.
%u: \citet{filippis2004}.
}

%\end{center}
%\end{sidewaystable*}
\end{deluxetable}

%----------------------------------------------------------------
\begin{deluxetable}{lccccccc}
\tabletypesize{\scriptsize}
%\tabletypesize{\normalsize} 
\rotate
\tablewidth{0pt}
\tablecaption{Observational Parameters
     \label{tab:obspars2}}
\tablehead{ \colhead{Cluster} & \colhead{redshift} & \colhead{exposure
time} & \colhead{average seeing}& \colhead{field of view} &
\colhead{native det. tresh.} & \colhead{corrected det. thresh} &
\colhead{back. accuracy}\\ \colhead{} & \colhead{} & \colhead{r, B/V hrs} &
\colhead{r,B/V arcsecs} & \colhead{\h Mpc} & \colhead{r,B/V\magarc} &
\colhead{r,B/V\magarc} & \colhead{r,B/V\magarc} }

\startdata
A4059 & 0.048  & 3.8,4.3  & 1.1,1.7  &  $1.2\times1.4$  & 24.9,25.8 & 26.25,26.04  & 27.7,29.9 \\
A3880 & 0.058  & 9.5,4.3  & 1.5,1.5  &  $1.7\times2.2$  & 25.1,25.2 & 26.22,26.04  & 28.0,29.2 \\
A2734 & 0.062  & 4.6,4.1  & 1.3,1.7  &  $1.7\times1.9$  & 24.8,25.2 & 26.21,26.05  & 28.7,29.3 \\
A2556 & 0.087  & 3.8,3.7  & 1.3,1.4  &  $2.0\times2.4$  & 24.7,25.4 & 26.13,26.11  & 27.7,29.4 \\
A4010 & 0.096  & 5.5,6.9  & 1.7,1.3  &  $2.3\times2.5$  & 25.0,25.7 & 26.10,26.09  & 28.4,29.9 \\ \\
\hline \\
A3888 & 0.151  & 6.3,4.0  & 1.1,0.9  &  $2.3\times2.2$  & 26.4,26.0 & 25.94,25.67  & 28.8,29.5 \\
A3984 & 0.181  & 6.8,4.3  & 1.0,1.0  &  $2.3\times2.5$  & 26.4,25.9 & 25.89,25.65  & 28.9,29.0 \\
A0141 & 0.23   & 6.8,4.0  & 0.9,1.0  &  $2.8\times2.7$  & 26.3,26.0 & 25.75,25.63  & 29.2,29.8 \\
AC118 & 0.31   & 6.0,3.0  & 1.4,1.0  &  $2.7\times2.9$  & 26.3,26.9 & 25.62,25.70  & 29.7,29.9 \\
AC114 & 0.31   & 5.5,4.3  & 1.1,1.8  &  $2.4\times2.5$  & 26.4,26.1 & 25.63,25.70  & 29.8,29.8 \\
\enddata

\tablecomments{ The first five clusters in the table were imaged with
the 1m Swope telescope in the $r-$ and $B-$bands.  The five clusters under
the middle line were imaged with the 2.5m DuPont telescope in the $r-$
and $V-$bands. Native detection threshold refers to the measured
detection threshold of the cluster at its appropriate redshift.
Corrected surface brightness detection threshold refers to the actual
detection threshold to which we mask at the redshift of each cluster.
This detection threshold has been surface brightness dimming and
k-corrected to a redshift of zero, as discussed in \S
\ref{galaxies2}.}
\end{deluxetable}

%-------------------------------------------------------------------
%-----------------------------------------------------------------

\newcommand\magarcs{mag arcsec$^{-2}$}

\clearpage
\begin{deluxetable}{lcccccccc}
\tabletypesize{\scriptsize}
%\tabletypesize{\normalsize} 
\tablewidth{0pt}
\tablecaption{Error Budget
     \label{tab:error2}}
\tablehead{
\colhead{Source} & \colhead{} & \colhead{} & \multicolumn{6}{c}{contribution to ICL uncertainty (\%)} \\
\colhead{} & \multicolumn{2}{c}{$1\sigma$ uncertainty} &\multicolumn{2}{c}{$\mu$(0\arcsec - 100\arcsec)} &\multicolumn{2}{c}{$\mu$(100\arcsec - 200\arcsec)} &\multicolumn{2}{c}{total ICL flux} \\
%\cline{2-3} \cline{4-5} \cline{6-7} \cline{8-9} \\
\colhead{} & \colhead{($V$)} & \colhead{($r$)} & \colhead{($V$)} & \colhead{($r$)} & \colhead{($V$)} & \colhead{($r$)} & \colhead{($V$)} & \colhead{($r$)}
}

\startdata

background level$^a$    & 29.5 \magarcs & 28.8 \magarcs & 14  & 18 & 39 & 45 & 24  & 31  \\
photometry              &  0.02 mag   & 0.03 mag    & 2   & 3  &  2 &  3 &  2  &  3  \\ 
masking$^b$             & \multicolumn{2}{c}{variation in mask area $\pm30$} & 5  & 5 & 14 & 19 & 9  & 12   \\
std.\ dev.\ in mean$^c$  &  32.7 \magarcs & 32.7 \magarcs & 3  & 2 & 2 & 1 & 3  & 1 \\ 
(total)                 & &                & 15 & 19 & 41 & 50 & 26 & 33\\ 
\\
\hline \\
cluster flux$^d$ &  16\% & 16\%  & \nodata & \nodata & \nodata & \nodata & \nodata  & \nodata  \\
\enddata

\tablecomments{ 
a: Large scale fluctuations in background level are measured
empirically and include instrumental calibration uncertainties 
as well as and true variations in background level (see \S \ref{noise2}).
b: Object masks were scaled by $\pm30\%$ in area to test the impact on
ICL measurement (see \S\ref{galaxies2}).
c: The statistical uncertainty in the mean surface brightness
of the ICL in each isophote.
d: Errors on the total cluster flux are based on errors in the fit to
the luminosity function (see \S \ref{member2}). 
}

\end{deluxetable}

%\clearpage
%-------------------------------------------------------------
%-------------------------------------------------------------
%----------------------------------------------------------------

%-----------------------------------------------------------------

%-------------------------------------------------------------

%-------------------------------------------------------------

%-------------------------------------------------------------

%-------------------------------------------------------------

%-------------------------------------------------------------

%-------------------------------------------------------------

%-------------------------------------------------------------

%-------------------------------------------------------------

%-------------------------------------------------------------

%-------------------------------------------------------------------

\clearpage

%-------------------------------------------------------------------
\clearpage

%-------------------------------------------------------------------
%-------------------------------------------------------------------

%-------------------------------------------------------------------

%-------------------------------------------------------------------

%-------------------------------------------------------------------

%-------------------------------------------------------------------

\clearpage
%-------------------------------------------------------------------

%-------------------------------------------------------------------
%-------------------------------------------------------------------

%-------------------------------------------------------------------

\clearpage

%-------------------------------------------------------------------
%-------------------------------------------------------------------
\clearpage

%-------------------------------------------------------------------------------
\appendix

\section{The Clusters}
In order of increasing redshift we discuss interesting characteristics
of the clusters and their ICL components.  Relevant papers are listed
in Table \ref{tab:characteristics2}.  Relevant figures are
\ref{fig:A4059} - \ref{fig:A118}.

\subsection{A4059}
\label{A4059}
A4059 is a richness class 1, Bautz Morgan type I cluster at a redshift
of 0.048.  There is a clear cD galaxy which is however offset from the
Abell center, likely due to the presence of at least two other bright
elliptical galaxies.  The cD galaxy is $0.91\pm.05$ magnitudes
brighter than the second ranked cluster galaxy.  The cD galaxy is at
the center of the Chandra and ASCA mass distributions.  Those
telescopes detect no hot gas around the other bright ellipticals.
This cluster shows interesting features in it's X--ray morphology.
There appear to be large bubbles, or cavities in the hot gas, which is
likely evidence of past radio galaxy interactions with the ICM
\citep{choi2004}.  As additional evidence of past activity in this
cluster, the cD galaxy contains a large dust lane
\citep{choi2004}. $M_{500}$ (the mass within the radius where the mean
mass density is equal to 500 times the critical density) is calculated
by \citet{reiprich2002} for A4059 to be $2.82\pm_{0.34}^{0.37} \times
10^{14} h_{70}^{-1}$ \msun.

The color magnitude diagram shows a very tight red sequence.
Membership information is taken from \citet{collins1995},
\citet{colless2001}, and \citet{smith2004}.  Using the CMD as an
indication of membership, we estimate the flux in cluster galaxies to
be $1.2 \pm .35\times10^{12}$\lsun in $r$ and
$4.2\pm1.3\times10^{11}$\lsun in $B$ inside of 0.65\h Mpc, which is
one quarter of the virial radius of this cluster.  In this particular
cluster, since the Abell center is not at the true cluster center, and
it is the nearest cluster in our sample, our image does not uniformly
cover the entire one quarter of the virial radius.  This estimate is
therefore below the true flux in galaxies because we are missing area
on the cluster.

Figure \ref{fig:A4059} shows the relevant plots for this cluster.
There is a strong ICL component ranging from 26 - 29 \magarc\ in $r$
centered on the cD galaxy.  The total flux in the ICL is
$3.4\pm1.7\times10^{11}$\lsun in $r$ and $1.2\pm.24\times10^{11}$\lsun
in $B$, which makes for ICL fractions of $22\pm12\%$ in $r$ and
$21\pm8\%$ in $B$.  The ICL has a flat color profile with $B-r \simeq
1.7\pm.08$, which is marginally bluer (0.2 magnitudes) than the RCS.
One of the two other bright ellipticals at 0.7\h Mpc from the center
has a diffuse component, the other bright elliptical is too close to a
saturated star to detect a diffuse component.

\subsection{A3880}
\label{A3880}

A3880 is a richness class 0, Bautz Morgan type II cluster at a
redshift of 0.058.  There is a clear cD galaxy in the center of this
cluster, which is $0.52\pm.05$ magnitudes brighter than the second
ranked galaxy. This cluster is detected in the ROSAT All Sky Survey,
however that survey is not deep enough to show us the shape of the
mass distribution.  \citet{girardi1998a} find a mass for this cluster
based on its velocity dispersion of $8.3_{-2.1}^{+2.8}\times10^{14}
h_{70}^{-1}\msun$.

The color magnitude diagram shows a clear red sequence.  There is
possibly another red sequence at lower redshift adding to the width of
the red sequence.  Membership information is provided by
\citet{collins1995}, \citet{colless2001}, and \citet{smith2004}.
Using the CMD as an indication of membership, we estimate the flux in
cluster galaxies to be $8.6\pm2.6\times10^{11}$\lsun in $r$ and
$3.8\pm1.1\times10^{11}$\lsun in $B$ inside of 0.62\h Mpc, which is one
quarter of the virial radius of this cluster.

Figure \ref{fig:A3880} shows the relevant plots for this cluster.
Unfortunately this cluster has larger illumination problems than the
other clusters which can be seen in the greyscale masked image.
Nonetheless, there is clearly an $r-$band ICL component, although the
$B-$band ICL is extremely faint.  The total flux in the ICL is
$1.4\pm2.3\times10^{11}$\lsun in $r$ and $4.4\pm1.5\times10^{10}$\lsun
in $B$, which makes for ICL fractions of $14\pm6\%$ in $r$ and
$10\pm6\%$ in $B$.  The ICL has a flat color profile with $B-r \simeq
2.4\pm1.1$, which is 0.8 magnitudes redder than the RCS.

\subsection{A2734}
\label{A2734}
A2734 is a richness class 1, Bautz Morgan type III cluster at a
redshift of 0.062.  The BCG by $0.51\pm.05$ magnitudes is in the center of
this cluster, however there are 2 other large elliptical galaxies
0.55\h Mpc and 0.85\h Mpc distant from the BCG. The X--ray gas does
confirm the BCG as being at the center of the mass distribution.
Those 2 other elliptical galaxies are not seen in the 44ks ASCA GIS
observation of this cluster, however they are confirmed members based
on spectroscopy \citep{collins1995,colless2001, smith2004}. $M_{500}$
is calculated by \citet{reiprich2002} for A2734 to be
$2.49\pm_{0.63}^{0.89} \times 10^{14} h_{70}^{-1}$ \msun.

The color magnitude diagram shows a clear red sequence, which includes
the 3 bright elliptical galaxies.  2df spectroscopy gives us roughly
80 galaxies in our field of view which we can use to estimate the
effectiveness of the biweight fit to the RCS in finding true cluster
members.  Of those galaxies with confirmed membership, 94\% are
determined members with this method, however 86\% of the confirmed
non-members are also considered members.  This is likely due to how
galaxies were selected for spectroscopy in the 2df catalog.  Using the
CMD as an indication of membership, we estimate the flux in cluster
galaxies to be $1.2\pm.36\times10^{12}$\lsun in $r$ and
$3.4\pm1.0\times10^{11}$\lsun in $B$ inside of 0.60\h Mpc, which is one
quarter of the virial radius of this cluster.

Figure \ref{fig:A2734} shows the relevant plots for this cluster.
There is a strong ICL component ranging from 26 - 29 \magarc\ in $r$
centered on the BCG.  The total flux in the ICL is
$2.8\pm.47\times10^{11}$\lsun in $r$ and $7.0\pm4.7\times10^{10}$\lsun
in $B$, which makes for ICL fractions of $19\pm6\%$ in $r$ and
$17\pm13\%$ in $B$.  The ICL has a flat to red-ward color profile with
$B-r \simeq 2.3\pm.03$, which is marginally redder than the RCS (0.3
magnitudes).  The cluster has a second diffuse light component around
one of the giant elliptical galaxies, .55 \h Mpc from the center of
the cD galaxy.  The third bright elliptical has a saturated star just
40\arcsec away, so we do not have a diffuse light map of that galaxy.

\subsection{A2556}
\label{A2556}
A2556 is a richness class 1, Bautz Morgan type II-III cluster at a
redshift of 0.087.  Despite the Bautz Morgan classification, this
cluster has a clear cD galaxy in the center of the X--ray distribution
which is $0.93\pm.05$ magnitudes brighter than any other galaxy in the
cluster.  The Chandra derived X--ray distribution is slightly
elongated toward the NE where a second cluster, A2554, resides, 1.4\h
Mpc from the center of A2556.  The cD galaxy of A2554 is just on the
edge of our images so we have no information about its low surface
brightness component. A2556 and A2554 are a part of the Aquarius
supercluster\citep{batuski1999}, so they clearly reside in an
overdense region of the universe.  Given an X--ray luminosity from
\citet{ebeling1996} and a velocity dispersion from
\citet{reimers1996}, we calculate the virial mass of A2556 to be
$2.5\pm1.1\times10^{15} h_{70}^{-1}\msun$.

The red sequence for this cluster is a bit wider than in other
clusters.  The one sigma width to a biweight fit is 0.38 magnitudes in
B-r which is approximately 30\% larger than in the rest of the low-z
sample.  This extra width is not caused by only a few galaxies,
instead the entire red sequence appears to be inflated.  This is
probably caused by the nearby A2554 which is at z=0.11
\citep{struble1999}.  This is close enough in redshift space that we
cannot separate out the 2 red sequences in our CMD.  We have roughly
30 redshifts for A2556 from \citet{smith2004}, \citet{caretta2004},
and \citet{batuski1999} which are also unable to differentiate between
the clusters.  Using the CMD as an indication of membership, we
estimate the flux in cluster galaxies to be $1.3\pm.38\times10^{12}$\lsun in
$r$ and $3.3\pm1.0\times10^{11}$\lsun in $B$ inside of 0.65\h Mpc, which is
one quarter of the virial radius of this cluster.

Figure \ref{fig:A2556} shows the relevant plots for this cluster.
There is an $r-$band ICL component ranging from 27 - 29 \magarc\ in
$r$ centered on the cD galaxy.  The $B-$band ICL is extremely faint,
barely above or detection threshold.  Although we were able to fit a
profile to the $B-$band diffuse light, all points on the medium sized
mask are below 29 \magarc. The total flux in the ICL is
$7.6\pm6.6\times10^{10}$\lsun in $r$ and $1.4\pm1.4\times10^{10}$\lsun
in $B$, which makes for ICL fractions of $6\pm5\%$ in $r$ and
$4\pm4\%$ in $B$.  Although Figure \ref{fig:A2556} shows a color
profile, we do not assume anything about the profile shape due to the
low SB level of the $B-$band.  We take the $B-r$ color from the
innermost point to be $2.1\pm0.4$, which is fully consistent with the
color of the RCS.

\subsection{A4010}
\label{A4010}

A4010 is a richness class 1, Bautz Morgan type I-II cluster at a
redshift of 0.096.  This cluster has a cD galaxy in the center of the
galaxy distribution, which is $0.7\pm.05$ magnitudes brighter than the second
ranked galaxy.  There is only ROSAT All Sky Survey data for this
cluster and no other sufficiently deep X--ray observations to show us
the shape of the mass distribution.  There are weak lensing maps which
put the center of mass of the cluster at the same position as the cD
galaxy, and elongated along the same position angle as the cD galaxy
\citep{cypriano2004}.  \citet{muriel2002} find a velocity dispersion
of $743\pm140$ for this cluster which is $15\%$ larger than found by
\citet{girardi1998a}, where those authors find a virial mass of
$3.8\pm_{1.2}^{1.6} \times 10^{14} h_{70}^{-1}$ \msun.

The color magnitude diagram for A4010 is typical among the sample with
a clear red sequence.  A few redshifts exist in the literature which
help define the red sequence \citep{collins1995, katgert1998}.  Using
the CMD as an indication of membership, we estimate the flux in
cluster galaxies to be $1.2\pm.4\times10^{12}$\lsun in $r$ and
$3.5\pm1.0\times10^{11}$\lsun in $B$ inside of 0.75\h Mpc, which is
one quarter of the virial radius of this cluster.

Figure \ref{fig:A4010} shows the relevant plots for this cluster.
There is an elongated ICL component ranging from 25.5 - 28 \magarc\ in
$r$ centered on the cD galaxy.  The total flux in the ICL is
$3.2\pm0.7\times10^{11}$\lsun in $r$ and $7.7\pm2.8\times10^{10}$\lsun
in $B$, which makes for ICL fractions of $21\pm8\%$ in $r$ and
$18\pm8\%$ in $B$.  The ICL has a significant red-ward trend in its
color profile with an average color of $B-r \simeq 2.1\pm0.1$, which
is marginally redder (0.2 magnitudes) than the RCS.

\subsection{A3888}
\label{A38882}
A3888 is discussed in great detail in paper I. In review, A3888 is a
richness class 2, Bautz Morgan type I-II cluster at a redshift of
0.151.  This cluster has no cD galaxy; instead the core is comprised
of 3 distinct sub-clumps of multiple galaxies each.  At least 2
galaxies in each of the subclumps are confirmed members based on
velocities \citep{teague1990, pimbblet2002}.  The brightest cluster
galaxy is only $0.12\pm.04$ magnitudes brighter than the second ranked
galaxy. XMM contours show an elongated distribution centered roughly
in the middle of the three clumps of galaxies.  \citet{reiprich2002}
estimate mass from the X--ray luminosity to be $M_{200}$ = 25.5$\pm
^{10.5} _{7.4}\times10^{14} h_{70}^{-1}$ \msun, where $r_{200}$ =
2.8$h^{-1}_{70}$Mpc.  This is consistent with the mass estimate from
the published velocity dispersion of $1102\pm^{137}_{107}$
\citep{girardi2001}.

There is a clear red sequence of galaxies in the CMD of A3888. Using
the CMD as an indication of membership, we estimate the flux in
cluster galaxies to be $3.0\pm0.9\times10^{12}$\lsun in $r$ and
$7.2\pm2.2\times10^{11}$\lsun in $B$ inside of 0.92\h Mpc.  We also
determine galaxy flux using the \citet{driver1998} luminosity
distribution, which is based on the statistical background subtraction
of non-cluster galaxies, to be $4.3\pm0.7\times10^{12}$\lsun in the
$r-$band and $3.4\pm0.6\times10^{12}$\lsun in $V$ .  The difference in
these two estimates is likely due to uncertainties in our membership
identification (of order $30\%$) and difference in detection
thresholds of the two surveys.

Figure \ref{fig:A3888} shows the relevant plots for this cluster.
There is a centralized ICL component ranging from 26 - 29 \magarc\ in
$r$ despite the fact that there is no cD galaxy.  The total flux in
the ICL is $4.4\pm2.1\times10^{11}$\lsun in $r$ and
$8.6\pm2.5\times10^{10}$\lsun in $B$, which makes for ICL fractions of
$13\pm5\%$ in $r$ and $11\pm3\%$ in $B$.  The ICL has a red color
profile with an average color of $V-r \simeq 0.5\pm0.1$, which is
marginally redder (0.2 magnitudes) than the RCS.  There is also a
diffuse light component surrounding a group of galaxies that is 1.4 \h
Mpc from the cluster center which totals $1.7\pm0.5\times10^{10}$\lsun
in V and $2.6\pm1.2\times10^{10}$\lsun in r and has a color consistent
with the main ICL component.

\subsection{A3984}
\label{A3984}

A3984 is an interesting richness class 2, Bautz Morgan type II-III
cluster at a redshift of 0.181.  There appear to be 2 centers of the
galaxy distribution.  One around the BCG, and one around a semi-circle
of $\sim5$ bright ellipticals which are 1\h Mpc north of the BCG.  The
BCG and at least one of the other bright ellipticals are at the same
redshift \citep{collins1995}. To determine if these 2 centers are part
of the same redshift structure, we split the image in half
perpendicular to the line bisecting the 2 regions, and plot the
cumulative distributions of $V-r$ galaxy colors.  A KS test reveals
that these 2 regions have an 89\% probability of being drawn from the
same distribution.  Without X--ray observations we do not know where
the mass in this cluster resides.  There is a weak lensing map of just
the northern region of the cluster which does show a centralized mass
distribution, but does not include the southern clump
\citep{cypriano2004}. The BCG is $0.57\pm.04$ magnitudes brighter than the
second ranked galaxy.  We use a velocity dispersion from the lensing
measurement to determine a mass of $31\pm10\times10^{14}
h_{70}^{-1}\msun$.

There is a clear red sequence of galaxies in the CMD of A3984. Using
the CMD as an indication of membership, we estimate the flux in
cluster galaxies to be $2.0\pm0.6\times10^{12}$\lsun in $r$ and
$4.4\pm1.3\times10^{11}$\lsun in $B$ inside of 0.87\h Mpc, which is
one quarter of the virial radius of this cluster.

Figure \ref{fig:A3984} shows the relevant plots for this cluster.
There are 2 clear groupings of diffuse light.  We can only fit a
profile to the ICL which is centered on the BCG.  We stop fitting that
profile before it extends into the other ICL group ($\sim 600kpc$) in
an attempt to keep the fluxes separate.  The total flux in the ICL is
$2.2\pm1.0\times10^{11}$\lsun in $r$ and $6.2\pm2.1\times10^{10}$\lsun
in $B$, which makes for ICL fractions of $10\pm6\%$ in $r$ and
$12\pm6\%$ in $B$.  The ICL becomes distinctly bluer with radius and
is bluer at all radii than the RCS with an average color of $V-r
\simeq -0.2\pm0.4$ (0.5 magnitudes bluer than the RCS).

\subsection{A0141}
\label{A0141}
A0141 is a richness class 3, Bautz Morgan type III cluster at a
redshift of 0.23.  True to its morphological type, this cluster has no
cD galaxy, instead it has 4 bright elliptical galaxies, each at the
center of a clump of galaxies, the brightest one of which is $0.42\pm.04$
magnitudes brighter than the second brightest.  The center of the
cluster, as defined by ASCA observations and a weak lensing map
\citep{dahle2002}, is near the northernmost clumps of galaxies.  The
distribution is clearly elongated north-south, it is therefore possible
that the other bright ellipticals are in-falling groups along a
filament.  $M_{200}$ from the lensing map is
$18.9\pm^{1.1}_{0.9}\times10^{14}$ \h \msun.

There is a clear red sequence of galaxies in the CMD of A0141. Using
the CMD as an indication of membership, we estimate the flux in
cluster galaxies to be $3.2\pm1.0\times10^{12}$\lsun in $r$ and
$5.4\pm1.6\times10^{11}$\lsun in $B$ inside of 0.94 \h Mpc, which is one
quarter of the virial radius of this cluster.

Figure \ref{fig:A141} shows the relevant plots for this cluster.
There are 3 clear groupings of diffuse light which do not have a
common center, although 1 of these ICL peaks does include 2 clumps of
galaxies.  We are unable to fit a single centralized profile to this
ICL as the three clumps are too far separated.  The total flux in the
ICL as measured in manually placed elliptical annuli is
$3.5\pm.9\times10^{11}$\lsun in $r$ and $3.4\pm1.1\times10^{10}$\lsun in $B$,
which makes for ICL fractions of $10\pm4\%$ in $r$ and $6\pm3\%$ in
$B$.  We estimate the color of the ICL to be $V-r \simeq 1.0\pm0.8$, which
is significantly redder (0.6 magnitudes) than the RCS.  We have no
color profile information.

\subsection{AC114}
\label{AC114}
AC114 (AS1077) is a richness class 2, Bautz Morgan type II-III cluster
at a redshift of 0.31.  The brightest galaxy is only $0.28\pm.04$
magnitudes brighter than the second ranked galaxy.  The galaxy
distribution is elongated southeast to northwest \citep{couch2001} as
is the Chandra derived X--ray distribution.  The X--ray gas shows a
very irregular morphology, with a soft X--ray tail stretching toward a
mass clump in the southeast which is also detected in a lensing map
\citep{defilippis2004,campusano2001}. The X--ray gas is roughly
centered on a bright elliptical galaxy, however the tail is an
indication of a recent interaction.  There is a clump of galaxies,
1.6\h Mpc northwest of the BCG, which looks like a group or cluster
with its own cD-like galaxy which is not targeted in either the X--ray
or lensing (strong) observations.  Only one of these galaxies has
redshifts in the literature, and it is a member of AC114.  Without
redshifts, we cannot know definitively if these galaxies are a part of
the same structure, however their location along the probable filament
might be evidence that they are part of the same velocity structure.
As this cluster is not in dynamical equilibrium, mass estimates from
the X--ray gas come from B-model fits to the surface brightness
distribution.  \citet{defilippis2004} find a mass within 1\h Mpc of
$4.5\pm1.1 \times 10^{14}$\h \msun.  A composite strong and weak
lensing analysis agree with the X--ray analysis within 500\h kpc, but
they do not extend out to larger radii \citep{campusano2001}.  Within
the virial radius, \citep{girardi2001} find a mass of
$26.3^{+8.2}_{-7.1} \times 10^{14}$\h \msun.

This cluster, in relation to lower-z clusters, is a prototypical
example of the Butcher-Oemler effect.  There is a higher fraction of
blue, late-type galaxies at this redshift, than in our lower-z
clusters, rising to 60\% outside of the core region \citep{couch1998}.
This is not only evidenced in the morphologies, but in the CMD, which
nicely shows these blue member galaxies.  We adopt the
\citet{andreon2005} luminosity function for this cluster based on an
extended likelihood distribution for background galaxies.  Integrating
the luminosity distribution from very dim dwarf galaxies
($M_{R}=-11.6$) to infinity gives a total luminosity for AC114 of
$1.5\pm0.2\times10^{12}$\lsun in $r$ and $1.9\pm1.2\times10^{11}$\lsun
in $B$ inside of 0.9\h Mpc, which is one quarter of the virial radius
of this cluster.  For the purpose of comparison with other clusters,
we adopt the cluster flux from the CMD, which gives
$1.8\pm0.5\times10^{12}$\lsun in $r$ and $2.3\pm0.7\times10^{11}$\lsun
in $B$ inside of one quarter of the virial radius of this cluster.
The differences in these estimates are likely due to uncertainties in
membership identification and differing detection thresholds of the
two surveys.

Figure \ref{fig:A114} shows the relevant plots for this cluster.
There is a centralized ICL component ranging from 27.5 - 29 \magarc\ in
$r$, in addition to a diffuse component around the group of galaxies
to the northwest of the BCG.  The total flux in the ICL is
$2.2\pm0.4\times10^{11}$\lsun in $r$ and $3.8\pm7.9\times10^{10}$\lsun in $B$,
which includes the flux from the group as measured in elliptical
annuli. The ICL fraction is $11\pm2\%$ in $r$ and $14\pm3\%$ in $B$.
The ICL has a flat color profile with $V-r \simeq 0.1\pm0.1$, which is
marginally bluer (0.4 magnitudes) than the RCS.

\subsection{AC118 (A2744)}
\label{AC118}

AC118 (A2744) is a richness class 3, Bautz Morgan type III cluster at
a redshift of 0.31.  This cluster has 2 main clumps of galaxies
separated by 1\h Mpc, with a third bright elliptical in a small group
which is 1.2\h Mpc distant from the center of the other clumps.  The
BCG is $0.23\pm.04$ magnitudes brighter than the second ranked galaxy.
The Chandra X--ray data suggests that there are probably 3 clusters here, at
least 2 of which are interacting.  The gas distribution, along with
abundance ratios, suggests that the third, smaller group might be the
core of one of the interacting clusters which has moved beyond the
scene of the interaction where the hot gas is detected.  From velocity
measurements \citet{girardi2001} also find 2 populations of galaxies
with distinctly different velocity dispersions.  The presence of a
large radio halo and radio relic are yet more evidence for dynamical
activity in this cluster \citep{govoni2001}.  Mass estimates for this
cluster range from $\sim3\times10^{13}\msun$ from X--ray data to
$\sim3\times10^{15}\msun$ from the velocity dispersion data.  This
cluster clearly violates assumptions of sphericity and hydrostatic
equilibrium, which is leading to the large variations.  The two
velocity dispersion peaks have a total mass of $38\pm37\times10^{14}$
\h \msun; we adopt this mass throughout the paper.

AC118, at the same redshift as AC114, also shows a significant
fraction of blue galaxies, which leads to a wider red cluster
sequence($1\sigma = 0.3$ magnitudes), than at lower redshifts. We
adopt the \citet{busarello2002} $R$ and $V-$band luminosity
distributions based on photometric redshifts and background counts
from a nearby, large area survey.  Integrating the luminosity
distribution from very dim dwarf galaxies ($M_{R}=-11.6$) to infinity
gives a total luminosity for AC118 of $4.5\pm.2\times10^{11}\lsun$
in $V$ and $4.2\pm.4\times10^{12}\lsun$ in the $r-$band inside of
$0.25r_{virial}$.  For the purpose of comparison with other clusters,
we adopt the cluster flux from the CMD, which gives
$5.4\pm1.6\times10^{11}$\lsun in $B$ and $4.4\pm0.1\times10^{12}$\lsun in $r$
inside of 0.94\h Mpc, which is one quarter of the virial radius of
this cluster.

Figure \ref{fig:A118} shows the relevant plots for this cluster.
There are at least two, if not three groupings of diffuse light which
do not have a common center.  The possible third is mostly obscured
behind the mask of a saturated star.  We are unable to fit a
centralized profile to this ICL.  The total flux in the ICL as
measured in manually placed elliptical annuli is
$7.0\pm1.0\times10^{11}$\lsun in $r$ and $6.7\pm1.7\times10^{10}$\lsun in $B$,
which makes for ICL fractions of $14\pm5\%$ in $r$ and $11\pm5\%$ in $B$.
We estimate the color of the ICL to be $V-r \simeq 1.0\pm0.8$, which is
significantly redder (0.6 magnitudes) than the RCS.  We have no color
profile information.

%---------------------------------------------------------------------------------

\end{document}